\newlength{\extralineskip}

  \documentstyle[12pt]{article}
\def\bea{\begin{eqnarray}}
\def\eea{\end{eqnarray}}

\newcounter{equnum}[section]
\def\theequnum{\thesection.\arabic{equnum}}
\newcommand{\beq}{$$ \refstepcounter{equnum}}
\newcommand{\eeq}{\eqno (\theequnum) $$}

\newcommand{\bd}{\begin{displaymath}}
\newcommand{\ed}{\end{displaymath}}

\font\twlmsy=msbm10 at 12pt
\font\sevenmsy=msbm8
\font\fivemsy=msbm6
\newfam\Bbbfam
\textfont\Bbbfam=\twlmsy
\scriptfont\Bbbfam=\sevenmsy
\scriptscriptfont\Bbbfam=\fivemsy
\def\Bbb{\fam\Bbbfam}

\makeatletter
\newdimen\normalarrayskip              
\newdimen\minarrayskip                 
\normalarrayskip\baselineskip
\minarrayskip\jot
\newif\ifold             \oldtrue            \def\new{\oldfalse}
\def\arraymode{\ifold\relax\else\displaystyle\fi} 
\def\@arrayskip{\ifold\baselineskip\z@\lineskip\z@
     \else
     \baselineskip\minarrayskip\lineskip2\minarrayskip\fi}
\def\@arrayclassz{\ifcase \@lastchclass \@acolampacol \or
\@ampacol \or \or \or \@addamp \or
   \@acolampacol \or \@firstampfalse \@acol \fi
\edef\@preamble{\@preamble
  \ifcase \@chnum
     \hfil$\relax\arraymode\@sharp$\hfil
     \or $\relax\arraymode\@sharp$\hfil
     \or \hfil$\relax\arraymode\@sharp$\fi}}
\def\@array[#1]#2{\setbox\@arstrutbox=\hbox{\vrule
     height\arraystretch \ht\strutbox
     depth\arraystretch \dp\strutbox
     width\z@}\@mkpream{#2}\edef\@preamble{\halign \noexpand\@halignto
\bgroup \tabskip\z@ \@arstrut \@preamble \tabskip\z@ \cr}%
\let\@startpbox\@@startpbox \let\@endpbox\@@endpbox
  \if #1t\vtop \else \if#1b\vbox \else \vcenter \fi\fi
  \bgroup \let\par\relax
  \let\@sharp##\let\protect\relax
  \@arrayskip\@preamble}
\makeatother

\def\sheet{{\mit\Sigma}}

\def\e{\, {\rm e}}
\def\MT{{{\cal M}_3}}
\def\M4{{{\cal M}_4}}

\begin{document}

\baselineskip=12pt

\begin{titlepage}
\begin{center}

\setcounter{footnote}0
\hfill OUTP-98-31P\\
\hfill hep-th/9804150\\
\hfill \\
\hfill Revised Version\\
\hfill \\
\hfill July 1998\\
\vspace{0.5in}

{\baselineskip=24pt

{\Large\bf String Holonomy and Extrinsic Geometry in \\Four-dimensional
Topological Gauge Theory}\\}

\vspace{0.5in}\normalsize

{\bf Richard J. Szabo}\footnote{\baselineskip=12pt Address after September 1,
1998: The Niels Bohr Institute, University of Copenhagen, Blegdamsvej 17,
DK-2100 Copenhagen \O, Denmark.}

\bigskip

\baselineskip=12pt

{\it Department of Physics -- Theoretical Physics\\ University of Oxford\\ 1
Keble Road, Oxford OX1 3NP, U.K.}\\ email: {\tt r.szabo1@physics.oxford.ac.uk}

\end{center}

\vskip 0.5 truein

\begin{abstract}

\baselineskip=12pt

The most general gauge-invariant marginal deformation of four-dimensional
abelian $BF$-type topological field theory is studied. It is shown that the
deformed quantum field theory is topological and that its observables compute,
in addition to the usual linking numbers, smooth intersection indices of
immersed surfaces which are related to the Euler and Chern characteristic
classes of their normal bundles in the underlying spacetime manifold. Canonical
quantization of the theory coupled to non-dynamical particle and string sources
is carried out in the Hamiltonian formalism and explicit solutions of the
Schr\"odinger equation are obtained. The wavefunctions carry a one-dimensional
unitary representation of the particle-string exchange holonomies and of
non-topological string-string intersection holonomies given by adiabatic limits
of the worldsheet Euler numbers. They also carry a multi-dimensional projective
representation of the deRham complex of the underlying spatial manifold and
define a generalization of the presentation of its motion group from Euclidean
space to an arbitrary 3-manifold. Some potential physical applications of the
topological field theory as a dual model for effective vortex strings are
discussed.

\end{abstract}

\vskip 0.5 truein

\leftline{{\sl PACS Numbers:} 11.15.-q , 11.25.-w , 71.10.p , 04.60.d}
\hfill 

\end{titlepage}

\clearpage\newpage

\setcounter{footnote}0

\baselineskip=18pt

\section{Introduction}

Topological quantum field theories \cite{schwarz,wittentqft} have many
applications in both physics and mathematics (see \cite{toprev} for a review).
They are characterized by the fact that their partition function and
observables are independent of the metric of the manifold on which they are
defined and therefore yield topological invariants of the underlying spacetime.
In some cases they constitute the effective quantum field theory of physical
models where topological phenomena, for example generalized Aharonov-Bohm
effects which arise from adiabatic transports of objects around one another,
play a significant role. They also provide interesting connections between
various seemingly disconnected branches of physics and mathematics, the classic
example being the interelation between knot theory, integrable models and
conformal field theory \cite{witten1}.

In this paper we will study a class of Schwarz-type topological gauge theories
\cite{schwarz,toprev}. One of the most widely studied examples of such theories
is given by the abelian Chern-Simons action \cite{schwarz,toprev,witten1}
\beq
S_{\rm CS}[A]=\int_\MT A\wedge F_A
\label{CSaction}\eeq
where $A$ is a $U(1)$ gauge connection of a complex line bundle over a
3-manifold $\MT$ and $F_A$ is its curvature. The quantum field theory defined
by (\ref{CSaction}) is strictly renormalizable and has appeared in a variety of
physical applications ranging from string theory \cite{carkogrev} to condensed
matter physics \cite{fqhe,ansc}. It provides a phenomenological realization of
anyons (see \cite{ansc,anrev} for reviews), i.e. particles in (2 +
1)-dimensions with fractional exchange statistics. Its observables yield
linking numbers of embedded curves in $\MT$ \cite{toprev,pol1} while those of
its nonabelian generalizations are related to polynomial invariants of knots
and links embedded in 3-manifolds \cite{witten1}.

A four-dimensional generalization of the topological field theory
(\ref{CSaction}) is defined by the abelian $BF$ action \cite{schwarz,toprev,BF}
\beq
S_{BF}[B,A]=\int_\M4 B\wedge F_A
\label{BFaction}\eeq
where $B$ is a 2-form field defined on a 4-manifold $\M4$ and $F_A$ is defined
as in (\ref{CSaction}). The corresponding partition function is related to the
Ray-Singer analytic torsion \cite{schwarz,toprev} which is a topological
invariant of $\M4$, and its observables compute the linking numbers of embedded
curves and surfaces in 4-manifolds \cite{toprev,BF}. It has been discussed in
connection with a wide variety of physical systems which involve vortex-like
configurations such as superconductors \cite{bal1}, cosmic strings
\cite{cosmic}, and axionic black holes \cite{axion}. Recent interest in the
nonabelian generalizations of this model has come from its role as a dual model
for quantum chromodynamics (QCD) whereby the exchange holonomies are relevant
to the quark confinement problem \cite{BFQCD}. The construction of explicit
physical states of the quantum field theory (\ref{BFaction}) which exhibit
particle-string ``fractional statistics" was carried out in \cite{bss}. The
physical relevance of these holonomies in string theory is discussed in
\cite{fracstring}, and for Nielsen-Olesen strings in abelian Higgs models in
\cite{ABphase}--\cite{topvortex}. In addition to their applications in
condensed matter physics, these latter models have also been relevant to
properties of the confining QCD string \cite{polQCD} and to the problem of
baryogenesis in electroweak theory \cite{electro}.

Unlike Chern-Simons theory, the quantum field theory defined by the $BF$ action
(\ref{BFaction}) is not stable under renormalization. Although it is well
established that $BF$ quantum field theories are ultraviolet finite in certain
gauges \cite{BFren}, it is still natural to examine the physical and
mathematical properties of the theory obtained by perturbing the action
(\ref{BFaction}) by gauge-invariant marginal (or irrelevant) operators. Such
deformations can be thought of as perturbing the quantum field theory
(\ref{BFaction}) to an isomorphic one in the associated moduli space. In the
following we will study the modification of this theory which is obtained by
renormalizing it via the addition of all truly marginal local operators to the
model (\ref{BFaction}). The renormalized model is a deformation of
(\ref{BFaction}) by a non-topological, explicitly metric-dependent counterterm,
but, as we demonstrate, the resulting action still defines a topological field
theory. This is similar to the usual situation in a topological gauge theory,
where the gauge-fixing couples the quantum action to the spacetime metric.
Nonetheless, the theory is topological since the energy-momentum tensor is
BRST-exact and therefore has vanishing matrix elements in physical states
\cite{toprev}, i.e. there are no classical propagating degrees of freedom. The
renormalized field theory will therefore work to describe the holonomy effects
which occur in adiabatic transport in a theory of point charges and strings
just as well as Chern-Simons theory describes anyons.

The crucial effect of the renormalized $BF$ field theory is that its
observables yield, in addition to the usual topological linking numbers of
embedded curves and surfaces in $\M4$, an effectively computable representation
of a novel intersection number of surfaces immersed in the manifold. This
quantity is only a smooth invariant of the surfaces and is related to the
geometry of their normal bundles in $\M4$. It can be expressed in terms of the
extrinsic geometry of the surfaces and also the Euler and Chern characteristic
classes of the normal bundles. We shall study this invariant in detail in both
an effective field theory formalism and in the context of canonical
quantization. In the former approach we will see that the effective action
contains the action for vortex strings with rigid extrinsic curvature term and
Polyakov $\theta$-term. It therefore serves as a dual model for the effective
field theory of the QCD string (and other vortex-like configurations), but in a
much different manner than the usual non-topological dual models of QCD do
\cite{BFQCD}. Some properties of these strings then become quite transparent
when viewed in this dual formalism. A similar sort of relationship was
established in \cite{lrs}, where the coupling of dynamical point particles to
nonabelian $BF$ gauge fields in two-dimensions was considered and related to
two-dimensional extended Poincar\'e gravity.

In the canonical formalism we will find an adiabatic limit of these
intersection numbers which in turn defines an adiabatic representation for the
Euler numbers of the associated normal bundles. The wavefunctions then carry,
in addition to the usual particle-string holonomies, a one-dimensional unitary
representation of a sort of non-topological string-string holonomy. These
holonomies give interesting representations of the extrinsic geometry of the
string worldsheets and could have applications in the aforementioned models. By
an explicit construction of the physical state wavefunctions, we show directly
that the physical Hilbert space is finite-dimensional and recover all of the
properties of the states of ordinary $BF$ theory \cite{bss}, but in a more
symmetric representation with respect to particle and string degrees of freedom
which also naturally provides a representation of the secondary gauge
constraints of the theory (\ref{BFaction}) \cite{toprev,BF}. These properties
include a multi-dimensional projective representation of the deRham complex of
the underlying spatial 3-manifold and of its associated motion group (the
generalization of the braid group in Chern-Simons field theory). The latter
feature was described briefly in \cite{bss} and here we shall expand somewhat
on the properties of this representation. In particular, we find that the $BF$
theory naturally defines the extension of the motion group presentation from
${\Bbb R}^3$ to an arbitrary 3-manifold.

The organisation of this paper is as follows. In section 2 we introduce the
perturbed $BF$ field theory and establish its topological properties. In
section 3 we examine the invariants represented by the effective field theory
when the gauge fields are coupled to non-dynamical particle and string sources,
and describe the potential physical applications of this effective model to
theories of vortex strings. In section 4 we describe the canonical structure
and reduced phase space of the theory, taking into careful account the
reducible gauge symmetries that $BF$ field theories possess. In section 5 we
construct the physical state wavefunctions which solve the gauge constraints
and the Schr\"odinger equation and develop the adiabatic representations of the
topological linking numbers, the extrinsic intersection indices and the Euler
numbers. In section 6 we describe in detail the transformation properties of
the physical states under gauge transformations and adiabatic transports. We
also explicitly construct a multi-dimensional representation of the motion
group which is valid for arbitrary 3-manifolds.

\section{Deformed $BF$ Field Theory}

Consider a real-valued 1-form field $A$ and a real-valued 2-form field
$B$ defined on a closed orientable four-dimensional spacetime manifold $\M4$
with metric $g$ of Minkowski signature. These forms can take values in some
flat vector bundle over $\M4$. They have the abelian gauge transforms
\beq
A\to A+\chi~~~~~~,~~~~~~B\to B+\xi
\label{gaugetr}\eeq
where $\chi$ is a closed 1-form and $\xi$ is a closed 2-form, $d\chi=d\xi=0$.
The $A$ field minimally couples to the particle current
\beq
j^\mu(x)=\sum_aq_a\int_{L_a}dl^\mu(r_a)~\delta^{(4)}(x,r_a(\tau))
\label{partcurr}\eeq
where
\beq
dl^\mu(r_a)=d\tau~{dr_a^\mu(\tau)\over d\tau}
\label{worldel}\eeq
is the differential particle worldline element and $r_a^\mu(\tau)$ is the
imbedding of the worldline $L_a$ of particle $a$ with charge $q_a$ in $\M4$. It
has dimension 3 and satisfies the continuity
equation $\partial_\mu j^\mu=0$ when the worldlines are closed. The $B$ field
couples minimally to the antisymmetric string current
\beq
\Sigma^{\mu\nu}(x)=\sum_b\phi_b\int_{{\mit\Sigma}_b}d\sigma^{\mu\nu}(X_b)
{}~\delta^{(4)}(x,X_b(\sigma))
\label{stringcurr}\eeq
where
\beq
d\sigma^{\mu\nu}(X_b)=d^2\sigma~\epsilon^{\alpha\beta}\,{\partial X_b^\mu(
\sigma)\over\partial\sigma^\alpha}{\partial X_b^\nu(\sigma)\over\partial
\sigma^\beta}
\label{strworldel}\eeq
is the differential string worldsheet area element
and $X_b^\mu(\sigma)$ is the imbedding of the worldsheet ${\mit\Sigma}_b$ of
string $b$ with electromagnetic flux $\phi_b$ in $\M4$. It has dimension 2
and obeys the conservation law $\partial_\mu\Sigma^{\mu\nu}=0$ when the
worldsheets are closed. We do not distinguish between the worldlines and
worldsheets and their imbeddings in $\M4$, which we assume to be disjoint,
$L_a\cap\sheet_b=\emptyset$.

A strictly renormalizable local field theory constructed from these fields must
include all gauge-invariant operators of dimension 4 or less. The action is
then\footnote{\baselineskip=12pt Here and in the following we will not write
explicit metric factors required to make all terms generally covariant.}
\beq\new{\begin{array}{c}
S[B,A;\Sigma,j]=\int_\M4
d^4x~\left(-{1\over4}F_{\mu\nu}F^{\mu\nu}+{k\over4\pi}\epsilon
^{\mu\nu\lambda\rho}B_{\mu\nu}\partial_\lambda A_\rho+{\theta\over4\pi}
\epsilon^{\mu\nu\lambda\rho}F_{\mu\nu}F_{\lambda\rho}\right.\\\left.+\,A_\mu
j^\mu+{1\over2}B_{\mu\nu}\Sigma^{\mu\nu}\right)\end{array}}
\label{renaction}\eeq
where $F=dA$ is the field strength of $A$ and we use the convention
$\epsilon^{0123}=+1$. The first term in (\ref{renaction}) is the usual Maxwell
term for the gauge field $A$ while the second term is the action
(\ref{BFaction}) of topological $BF$ theory. The third term is the usual
topological action of four-dimensional Yang-Mills theory and is a total
derivative. It is non-trivial only on spacetimes with non-contractible loops
and it does not appear in the classical equations of motion.

The field equations which follow from the action (\ref{renaction}) are
\beq
{k\over4\pi}\epsilon^{\mu\nu\lambda\rho}F_{\lambda\rho}+\Sigma^{\mu\nu}=0
{}~~~~~~,~~~~~~-\partial_\nu
F^{\mu\nu}+{k\over4\pi}\epsilon^{\mu\nu\lambda\rho}\partial
_\nu B_{\lambda\rho}+j^\mu=0
\label{fieldeq}\eeq
The first field equation in (\ref{fieldeq}) confines electromagnetic flux to
the string worldsheets and gives the analog of the Meissner effect in a BCS
superconductor. The solutions of these field equations in a covariant gauge
$\partial^\mu A_\mu=\partial^\mu B_{\mu\nu}=0$ are
\beq
A_\mu=-{2\pi\over k}\epsilon_{\mu\nu\lambda\rho}{\partial^\nu\over\Box}
\Sigma^{\lambda\rho}~~~~~~,~~~~~~B_{\mu\nu}={16\pi^2\over
k^2}\Sigma_{\mu\nu}-{4\pi
\over k}\epsilon_{\mu\nu\lambda\rho}{\partial^\lambda\over\Box}j^\rho
\label{fieldeqsol}\eeq
where in this section and the next we shall ignore harmonic zero modes on
$\M4$. Substituting (\ref{fieldeqsol}) into (\ref{renaction}), we see that the
effective classical action is
\beq
\Gamma[\Sigma,j]=\int_\M4 d^4x~\left({4\pi^2\over
k^2}\Sigma^{\mu\nu}\Sigma_{\mu
\nu}-{4\pi\theta\over k^2}\epsilon_{\mu\nu\lambda\rho}\Sigma^{\mu\nu}\Sigma
^{\lambda\rho}-{2\pi\over k}j^\mu\epsilon_{\mu\nu\lambda\rho}{\partial^\nu
\over\Box}\Sigma^{\lambda\rho}\right)
\label{effaction}\eeq

The effective action (\ref{effaction}) shows that there are no propagating
degrees of freedom in this theory and therefore the action (\ref{renaction})
essentially defines a topological field theory. Unlike other topological field
theories, however, the action (\ref{renaction}) explicitly couples to the
spacetime metric. Nonetheless, from (\ref{fieldeq}) and (\ref{gaugetr}) it
follows that, in the source-free case, the space of classical solutions of the
field theory (\ref{renaction}) is the finite-dimensional vector space
$H^1(\M4)\oplus H^2(\M4)$ which coincides with that of the $BF$ field theory
(\ref{BFaction}) ($H^k(\M4)$ is the deRham cohomology group of $k$-forms on
$\M4$ with values in a flat vector bundle). The trace-less gauge-invariant
symmetric energy-momentum tensor is
\beq
T_{\mu\nu}\equiv\frac{\delta S}{\delta
g^{\mu\nu}}\biggm|_{j=\Sigma=0}=\frac14g_{\mu\nu}\,F_{\lambda\rho}F^{\lambda
\rho}-F_{\mu\lambda}F^{~\lambda}_\nu
\label{emtensor}\eeq
and it vanishes when restricted to flat gauge connections (i.e. classical field
configurations). It is also possible to prove that the quantum field theory
defined by (\ref{renaction}) is topological. For this, we regard the $F^2$
terms in the source-free action (\ref{renaction}) as a deformation $\Delta$ of
the topological $BF$ theory (\ref{BFaction}),
\beq
S[B,A;0,0]=(k/4\pi)\,S_{BF}[B,A]+\Delta[A]
\label{BFdeform}\eeq
Consider the local gauge symmetries (\ref{gaugetr}) which are parametrized by
smooth functions $\chi'$ with $\chi=d\chi'$ and 1-forms $\xi'$ with
$\xi=d\xi'$. The gauge transformations are then
\beq\new{\begin{array}{lll}
\delta_{\chi'}A=d\chi'~~~~~~&,&~~~~~~\delta_{\chi'}B=0\\\delta_{\xi'}A=0~~~~~~&,
&~~~~~~\delta_{\xi'}B=d\xi'\end{array}}
\label{gaugesyms}\eeq
For the $BF$ field theory (\ref{BFaction}), this symmetry is off-shell
reducible, and, therefore, in addition to the usual ghost fields required for
gauge-fixing, there are ghost-for-ghost fields due to the secondary gauge
invariance
\beq
\delta_{\chi''}\xi'=d\chi''
\label{secondgaugesym}\eeq

To gauge-fix the invariances (\ref{gaugesyms}), we introduce as usual a
Faddeev-Popov 0-form ghost field $c$ of ghost number 1 for $A$, and a 1-form
ghost field $c_1$ of ghost number 1 for $B$. The gauge-fixing of
(\ref{secondgaugesym}) is then achieved by introducing an additional 0-form
ghost $c_0$ with grading 2. One can now introduce the usual St\"uckelberg
fields and Faddeev-Popov anti-ghost fields to write down the gauge-fixed
quantum action corresponding to (\ref{BFaction}) \cite{toprev,BF}. Here we
shall work instead in the Batalin-Vilkovisky antifield-antibracket formalism
\cite{toprev,BVBF,BFYMlong}. For each set of fields $(A,c)$ and $(B,c_1,c_0)$
we introduce the corresponding set of antifields $(A^*,c^*)$ and
$(B^*,c_3^*,c_4^*)$ of dual form degrees $(3,4)$ and $(2,3,4)$ and ghost
numbers $(-1,-2)$ and $(-1,-2,-3)$, respectively. The fields and antifields
together generate the BRST symmetry of the gauge-fixed topological field
theory. The classical theory is recovered by setting all antifields to 0. The
off-shell nilpotent BRST algebra is represented as \cite{BVBF,BFYMlong}
\beq\new{\begin{array}{rrllrrl}
\left\{{\cal Q}\,,\,c_4^*\right\}_+&=&-dc_3^*~~~~~~& &~~~~~~\left\{{\cal
Q}\,,\,c^*\right\}_+&=&-dA^*\\\left\{{\cal
Q}\,,\,c_3^*\right\}_+&=&-dB^*~~~~~~& &~~~~~~\left\{{\cal
Q}\,,\,A^*\right\}_+&=&-dB\\\left\{{\cal Q}\,,\,B^*\right\}_+&=&-dA~~~~~~&
&~~~~~~\left\{{\cal Q}\,,\,B\right\}_+&=&-dc_1\\\left\{{\cal
Q}\,,\,A\right\}_+&=&-dc~~~~~~& &~~~~~~\left\{{\cal
Q}\,,\,c_1\right\}_+&=&-dc_0\\\left\{{\cal Q}\,,\,c\right\}_+&=&0~~~~~~&
&~~~~~~\left\{{\cal Q}\,,\,c_0\right\}_+&=&0\end{array}}
\label{BRSTalg}\eeq
where
\beq
{\cal Q}=\int_\M4\Bigl(B\wedge dA+A^*\wedge dc+B^*\wedge dc_1+c_3^*\wedge
dc_0\Bigr)
\label{BRSTcharge}\eeq
is the metric-independent BRST supercharge in the field-antifield
representation. The bracket in (\ref{BRSTalg}) denotes the graded antibracket
acting in the $\Bbb Z$-graded algebra of functionals of the fields and
antifields.

{}From (\ref{BRSTalg}) it follows that the deformation in (\ref{BFdeform}) can
be written as
\beq
\Delta[A]=\int_\M4\Bigl\{{\cal Q}\,,\,\Psi\Bigr\}_+
\label{BRSTexactdef}\eeq
where
\beq
\Psi=B^*\wedge\left(\mbox{$-\frac14$}\,\star
F+\mbox{$\frac\theta{4\pi}$}\,F\right)
\label{gaugeferm}\eeq
is a gauge fermion field. Here $\star$ denotes the Hodge duality operator which
is constructed from the metric of $\M4$ and acts on the exterior algebra
$\Lambda\M4$. The BRST-exact representation of the deformation immediately
implies that (\ref{renaction}) defines a topological quantum field theory. To
see this, we consider the partition function $\cal Z$ which, ignoring the
irrelevant gauge-fixing terms for the present
discussion\footnote{\baselineskip=12pt The gauge-fixing terms for the
renormalized field theory (\ref{renaction}) will be the same as those of the
pure $BF$ theory (\ref{BFaction}). Consequently, the full gauge-fixed quantum
field theory will also be topological in the same way that ordinary $BF$ field
theory is. Furthermore, the associated Faddeev-Popov determinants and
St\"uckelberg Jacobians will cancel each other in the usual way
\cite{toprev}.}, is given symbolically by the path integral
\beq
{\cal Z}=\int DA~DB~\e^{iS[B,A;0,0]}
\label{partfndef}\eeq
Since all the metric dependence of the action (\ref{renaction}) lies in the
gauge fermion field (\ref{gaugeferm}), the variation of (\ref{partfndef}) with
respect to the metric $g$ of $\M4$ is given by
\beq
-i\frac{\delta{\cal Z}}{\delta g^{\mu\nu}}=\Bigl\langle
T_{\mu\nu}\Bigr\rangle=\left\langle\left\{{\cal Q}\,,\,\frac{\delta\Psi}{\delta
g^{\mu\nu}}\right\}_+\right\rangle
\label{partfnvarg}\eeq
where the averages denote vacuum expectation values in the source-free theory
(\ref{renaction}). By gauge invariance, the BRST charge $\cal Q$ annihilates
the vacuum of the quantum field theory and (\ref{partfnvarg}) vanishes. The
partition function (\ref{partfndef}) is thus formally independent of the metric
of $\M4$ and there are no local degrees of freedom in this model. The same
property can be derived for the path integral representing the vacuum
expectation value of any metric-independent gauge-invariant operator ${\cal
O}[B,A]$ (so that $\{{\cal Q},{\cal O}\}_+=0$) of the source-free field theory
(\ref{renaction}).

The topological gauge theory (\ref{renaction}) will be henceforth refered to as
``deformed $BF$ theory". From (\ref{BRSTexactdef}) and (\ref{gaugeferm}) we see
that deformed $BF$ theory is a perturbation of ordinary $BF$ theory in which
the deformation can be considered as a natural metric-dependent $BF$ term
involving the Batalin-Vilkovisky antifield of the field $B$. The gauge fermion
field $\Psi$ thus depends on fields of the non-minimal sector of this
formalism, in that $B^*$ is not a gauge-fixing field but rather an auxilliary
field required to close the BRST algebra (\ref{BRSTalg}). This is a significant
difference between the counterterms in (\ref{renaction}) and the usual
metric-dependent gauge-fixing counterterms that are added to topological field
theory actions. These properties also distinguish the model (\ref{renaction})
from the deformations of $BF$ theory which are dual models for Yang-Mills
theory \cite{BFYMlong,BFYM}. These models perturb the action (\ref{BFaction})
by (gauge non-invariant) $B^2$ terms, and integrating over the $B$ field in the
deformed quantum field theory yields exactly the Yang-Mills model.

In the physical sector of the quantum field theory (localized about the
classical flat gauge field configurations), by definition $B^*=0$, and thus the
partition function of the model (\ref{renaction}) will represent the same
topological invariant (the Ray-Singer analytic torsion) of $\M4$ as that of the
unperturbed $BF$ field theory (\ref{BFaction}) \cite{schwarz,toprev,BF}. This
will not be true, however, of its observables. In the following we shall be
interested in the type of holonomy effects that the topological field theory
(\ref{renaction}) describes, i.e. the holonomy which occurs in adiabatic
transport in a theory of point charges and strings where all other degrees of
freedom are heavy. This has been extensively studied for the conventional
$B\wedge F$ action \cite{bss} to which we shall compare the properties of the
modified theory.

\section{Holonomy Operators and Effective Field Theory}

In this section we shall examine the topological and geometrical quantities
represented by the effective action
\beq
\Gamma[\Sigma,j]=-i\log\int
DA~DB~\e^{iS[B,A;\Sigma,j]}=-i\log\left\langle\prod_{a,b}W[L_a]\,W[\sheet_b]
\right\rangle
\label{effactionpath}\eeq
which corresponds to the expectation values of the Wilson line and surface
operators $W[L_a]=\exp iq_a\int_{L_a}A$ and
$W[\sheet_b]=\exp\frac{i\phi_b}2\int_{\sheet_b}B$ for the particles and strings
in the pure gauge part of (\ref{renaction}). It therefore represents the
generic gauge- and topologically-invariant observables of the quantum field
theory. Although in this paper we shall be primarily interested in the
canonical structure of the quantum field theory (\ref{renaction}), it is
instructive to first examine what sort of invariants the theory will represent
in a covariant framework. From (\ref{effaction}) we see that the effective
action consists of three separate terms, the first two being string-string
interactions and the last one particle-string interactions. It represents the
holonomy phase factors of the covariant particle-string composite states. We
shall see that these phase factors can all be written in terms of the extrinsic
geometry of the strings, and topological and geometrical intersection indices.

\subsection{Topological Linking Numbers}

The particle-string interaction term in (\ref{effaction}) is the effective
action of ordinary $BF$ field theory and is a topological linking number of the
string and particle trajectories in $\M4$. To see this, we use the
explicit representations (\ref{partcurr}) and (\ref{stringcurr}) for the
sources, the continuity equations and Stokes' theorem to write it as
\beq\new{\begin{array}{lll}
S_L[\Sigma,j]&\equiv&-{2\pi\over k}\int_\M4
d^4x~j^\mu(x)\epsilon_{\mu\nu\lambda
\rho}{\partial^\nu\over\Box}\Sigma^{\lambda\rho}(x)\cr&
=&-{2\pi\over k}\int\!\!\!\int_\M4 d^4x~d^4y~\epsilon_{\mu\nu\lambda
\rho}j^\mu(x)\,(x|\partial^\nu/\Box|y)\,\Sigma^{\lambda\rho}(y)
\\&=&-{2\pi\over k}\sum_{a,b}q_a\phi_b\,I(L_a,{\mit\Sigma}_b)\end{array}}
\label{linkaction}\eeq
where
\beq
I(L_a,{\mit\Sigma}_b)=\int_{B_x({\mit\Sigma}_b)}\int_{L_a}\delta^{(1,3)}
(r_a(\tau),x)=-\int_{D_x(L_a)}\int_{{\mit\Sigma}_b}\delta^{(2,2)}(X_b(\sigma),x)
\label{linknum}\eeq
is the covariant linking number of the worldline $L_a$ with the worldsheet
${\mit\Sigma}_b$ in $\M4$. Here $B({\mit\Sigma}_b)$ is a volume bounded by the
surface ${\mit\Sigma}_b$ and $D(L_a)$ is a disk whose boundary is the contour
$L_a$ in $\M4$. $\delta^{(1,3)}(x,y)$ is the Dirac delta-function in the
exterior algebra $\Lambda^1(\M4(x))\otimes\Lambda^3(\M4(y))$ with the property
that $\int_{\M4(y)}\delta^{(1,3)}(x,y)\wedge\alpha(y)=\alpha(x)$ for any 1-form
$\alpha(x)\in\Lambda^1(\M4(x))$. Its contour integral over $L_a$ defines
the deRham current 3-form $\triangle_{L_a}$ \cite{botttu} which is the
Poincar\'e dual of $L_a$ in $\M4$, i.e.
$\int_\M4\triangle_{L_a}\wedge\alpha=\int_{L_a}\alpha$. Likewise
$\delta^{(2,2)}(x,y)$ is the Dirac delta-function in the space
$\Lambda^2(\M4(x))\otimes\Lambda^2(\M4(y))$ so that
$\int_{\M4(y)}\delta^{(2,2)}(x,y)\wedge\beta(y)=\beta(x)$ for any 2-form
$\beta(x)\in\Lambda^2(\M4(x))$. Its surface integral over ${\mit\Sigma}_b$
gives the deRham current 2-form $\triangle_{{\mit\Sigma}_b}$ which is the
Poincar\'e dual of ${\mit\Sigma}_b$ in $\M4$, i.e.
$\int_\M4\triangle_{{\mit\Sigma}_b}\wedge\beta=\int_{{\mit\Sigma}_b}\beta$. The
covariant linking number (\ref{linknum}) can therefore be alternatively written
in terms of cohomology classes dual to the worldline and worldsheet homology
classes as
\beq
I(L_a,\sheet_b)=\int_{B(\sheet_b)}\triangle_{L_a}=-\int_{D(L_a)}
\triangle_{\sheet_b}
\label{IDRrep}\eeq
When $\M4={\Bbb R}^1\times{\Bbb R}^3$, (\ref{linknum}) becomes the standard
Gauss linking integral in four-dimensions \cite{bss}.

The quantity (\ref{linknum}) counts the number of times particle $a$ and string
$b$ link themselves in $\M4$ and is a topological invariant of the particle and
string trajectories. This linking number is the signed intersection number of
the line $L_a$ with the volume $B({\mit\Sigma_b})$, or equivalently
of the surface ${\mit\Sigma}_b$ with the disk $D(L_a)$, in $\M4$
\footnote{\baselineskip=12pt Generically, in $d$ dimensions, a $p$-surface and
a
$(d-p)$-surface intersect transversally at distinct isolated points.}
\beq
I(L_a,{\mit\Sigma}_b)=\sum_{p\in L_a\cap B({\mit\Sigma}_b)}{\rm sgn}(p)
=-\sum_{p\in{\mit\Sigma_b}\cap D(L_a)}{\rm sgn}(p)
\label{intnum}\eeq
where ${\rm sgn}(p)=\pm1$ according to whether or not the orientation at
the intersection point $p$ coincides with that of $\M4$. In the path integral
quantization of the field theory, it is the linking term (\ref{linkaction})
that endows the effective particle-string composite states with fractional
exchange statistics. Here the statistics parameter is ${k\over2\pi}$.

\subsection{Extrinsic Geometry of Strings}

The two local string-string interaction terms in the effective action
(\ref{effaction}) can be written in terms of local intersection indices and the
extrinsic geometry of the string worldsheets in $\M4$. Substituting into the
first term of (\ref{effaction}) the explicit form (\ref{stringcurr}) of the
string current and integrating over $x$, we see that it can be written as
\beq
S_E[\Sigma]\equiv{4\pi^2\over k^2}\int_\M4
d^4x~\Sigma^{\mu\nu}(x)\Sigma_{\mu\nu}(x)
=\sum_{b,b'}S_E^{(bb')}\label{sesum}\eeq
where
\beq
S_E^{(bb')}={4\pi^2\over k^2}\phi_b\phi_{b'}\int_{\sheet_b}d^2\sigma~
\sqrt{\eta_b(\sigma)}\int_{\sheet_{b'}}d^2\sigma'~\sqrt{\eta_{b'}(
\sigma')}~t_{b,\mu\nu}(\sigma)t^{\mu\nu}_{b'}(\sigma')\,\delta^{(4)}(X_
{b'}(\sigma'),X_b(\sigma))\label{sebbp}\eeq
Here
\beq
t_b^{\mu\nu}(\sigma)={1\over\sqrt{\eta_b(\sigma)}}\,\epsilon^{\alpha\beta}
{\partial X_b^\mu(\sigma)\over\partial\sigma^\alpha}{\partial X_b^\nu(\sigma)
\over\partial\sigma^\beta}\label{areaelt}\eeq
is the antisymmetric local area element of the surface $\sheet_b$ obeying the
identities
\beq
t_{b,\mu\nu}(\sigma)t_b^{\mu\nu}(\sigma)=2~~~~~~,~~~~~~\epsilon^{\mu\nu
\lambda\rho}t_{b,\mu\nu}(\sigma)t_{b,\lambda\rho}(\sigma)=0
\label{areaeltid}\eeq
and
\beq
\eta_{b,\alpha\beta}(\sigma)={\partial X_b^\mu(\sigma)\over\partial\sigma^
\alpha}{\partial X_{b,\mu}(\sigma)\over\partial\sigma^\beta}
\label{fundform1}\eeq
is the induced metric on the string worldsheet formed by the tangent vectors
${\partial X_b^\mu(\sigma)\over\partial\sigma^\alpha}$ to $\sheet_b\subset\M4$.
It has matrix inverse $\eta_b^{\alpha\beta}(\sigma)$ and determinant
$\eta_b(\sigma)$.

The integrals in (\ref{sebbp}) for $b=b'$ are localized onto the subset ${\cal
C}_b$ of the manifold $\sheet_b(\sigma)\otimes\sheet_b(\sigma')$ of points
$\{\sigma,\sigma'\}$ for which $X_b^\mu(\sigma)=X_b^\mu(\sigma')$. It can be
decomposed into two disjoint subsets, ${\cal C}_b={\cal E}_b\amalg{\cal N}_b$,
where ${\cal E}_b=\left\{\{\sigma,
\sigma'\}~|~X_b(\sigma)=X_b(\sigma')\iff\sigma=\sigma'\right\}$ and ${\cal
N}_b=
\left\{\{\sigma,\sigma'\}~|~X_b(\sigma)=X_b(\sigma'),\sigma\neq\sigma'\right\}$.
The integrals (\ref{sebbp}) for $b=b'$ are then the sum of the contributions
$S_E^{(bb)}=S_E({\cal E}_b)+S_E({\cal N}_b)$ from these two disjoint subsets.
In this subsection we shall examine the contribution from the subset ${\cal
E}_b$ on which the string functions $X_b$ are embeddings. The contribution from
${\cal N}_b$, which corresponds to twists or self-intersections of the immersed
surface $\sheet_b$, along with the $b\neq b'$ terms in (\ref{sebbp}) will be
described in the next subsection.

To describe these terms, let us briefly recall a few facts concerning the
extrinsic geometry of embedded surfaces. The embedding of $\sheet_b$ in $\M4$
can be used to define the tangent bundle $T\sheet_b$ over $\sheet_b$. The fibre
over a point $\sigma\in\sheet_b$ is the space of tangent vectors at the point
$X_b^\mu(\sigma)\in\M4$ (i.e. linear combinations of ${\partial X_b^\mu(
\sigma)\over\partial\sigma^\alpha}$). The associated normal bundle $N\sheet_b$
can be similarly defined, with fibre over the point $\sigma\in\sheet_b$ the
space of vectors orthogonal to the tangent vectors at $X_b^\mu(\sigma)$. The
normal fibres are spanned by normal vectors $n^\mu_{b,\ell}$, $\ell=1,2$, which
satisfy
\beq
n^\mu_{b,\ell}n_{b,\ell';\mu}=\delta_{\ell\ell'}~~~~~~,~~~~~~n_{b,\ell;\mu}
{\partial X_b^\mu\over\partial\sigma^\alpha}=0
\label{normvec}\eeq
The extrinsic curvature $K^{\alpha\beta}_{b,\ell}$ of the normal bundle is
defined by decomposing the Hessian of the embedding $X_b$ as an endomorphism
over the vector space $T\sheet_b\oplus N\sheet_b\cong T\M4$,
\beq
{\partial^2X_b^\mu(\sigma)\over\partial\sigma^\alpha\partial\sigma^\beta}
=\Gamma_{b,\alpha\beta}^\gamma{\partial X_b^\mu(\sigma)\over\partial\sigma
^\gamma}+K_{b,\ell;\alpha\beta}\,n^\mu_{b,\ell}
\label{embhess}\eeq
where $\Gamma_{b,\alpha\beta}^{\gamma}$ is the Christoffel connection on
$T\sheet_b$. Note that the intrinsic Gaussian curvature $R_b$ of the worldsheet
$\sheet_b$ is related to the curvature of the induced metric (\ref{fundform1})
by
\beq
R_b=\left(K^\alpha_{b,\ell;\alpha}\right)^2-K^\alpha_{b,\ell;\beta}
K^\beta_{b,\ell;\alpha}
\label{extcurv}\eeq

To evaluate the delta-function in (\ref{sebbp}) over $\sheet_b$, we note that
it is determined by the topology of the normal bundle $N\sheet_b$, because it
can defined as the limit of non-singular forms with shrinking supports in the
neighbourhood of $\sheet_b$, which in turn can be approximated by the zero
section of $N\sheet_b$. Thus we write
\beq
\delta^{(4)}(x,y)=\lim_{\Lambda\to\infty}\psi_\Lambda(x,y)
\label{regdelta}\eeq
where $\{\psi_\Lambda(x,y)\}_{\Lambda\in{\Bbb R}^+}$ is a one-parameter family
of smoothly supported functions near $x=y$ with
$\int_\M4\star\,\psi_\Lambda=1$. Working in Gaussian normal coordinates in the
transverse space of $\sheet_b$ (i.e. in Cartesian coordinates in the fibre of
$N\sheet_b$), we can choose
$\psi_\Lambda(x,y)=C_\Lambda\e^{-\Lambda^2\|x-y\|^2}$ where $\|x\|$ denotes the
geodesic length of $x$ in $\M4$. If $\M4$ is an open infinite manifold, then
the normalization constant $C_\Lambda$ diverges as $\Lambda^4$ for
$\Lambda\to\infty$. In general we are therefore concerned with the evaluation
of generic integrals of the form
\beq
{\cal Z}_b(\sigma)=\lim_{\Lambda\to\infty}C_\Lambda\int_{\sheet_b}
d^2\sigma'~\sqrt{\eta_b(\sigma')}~{\cal K}(\sigma,\sigma')~\e^{-\Lambda^2
\|X_b(\sigma')-X_b(\sigma)\|^2}
\label{genintev}\eeq
where ${\cal K}(\sigma,\sigma')$ is a local integration kernel on
$\sheet_b(\sigma)\otimes\sheet_b(\sigma')$. Since for
$\{\sigma,\sigma'\}\in{\cal E}_b$, $X_b^\mu(\sigma)=X_b^\mu(\sigma')$
is equivalent to $\sigma=\sigma'$, the geodesic function appearing in the
argument of the exponential in (\ref{genintev}) is a Morse function of
$\sigma'\in\sheet_b$ with global minimum $0$ at $\sigma'=\sigma$. We can
therefore apply the stationary-phase approximation to evaluate the integral
(\ref{genintev}) which yields the standard expansion \cite{hormander}
\beq\new{\begin{array}{lll}
{\cal Z}_b(\sigma)&=&\lim_{\Lambda\to\infty}C_\Lambda\Biggm
\{\left(-{2\pi\over\Lambda^2}\right){\det}^{-1/2}{\cal H}_b(\sigma)\cr&
&\times\sum_{\ell=0,1}\left({-1\over2\Lambda^2}\right)^\ell\left[{\cal H}_
b^{-1}(\sigma)^{\alpha\beta}{\partial^2\over\partial\sigma'^\alpha
\partial\sigma'^\beta}\right]^\ell{\cal K}(\sigma,\sigma')\biggm|_{\sigma'=
\sigma}+{\cal O}(\Lambda^{-6})\Biggm\}\end{array}}
\label{zbstat}\eeq
where ${\cal H}_b(\sigma)$ is the Hessian of the exponential argument in
(\ref{genintev}) at $\sigma'=\sigma$. One easily finds ${\cal
H}_{b,\alpha\beta}=\eta_{b,\alpha\beta}$ and thus
\beq
{\cal
Z}_b(\sigma)=2\pi\lim_{\Lambda\to\infty}\left(\frac{C_\Lambda}{\Lambda^2}{{\cal
K}(\sigma,\sigma)\over\sqrt{
\eta_b(\sigma)}}-\frac{C_\Lambda}{2\Lambda^4}{\eta_b^{\alpha\beta}(\sigma)
\over\sqrt{\eta_b(\sigma)}}{\partial^2{\cal K}(\sigma,\sigma')\over\partial
\sigma^\alpha\partial\sigma'^\beta}\biggm|_{\sigma'=\sigma}\right)
\label{zbgen}\eeq

Substituting ${\cal K}(\sigma,\sigma')=t_{b,\mu\nu}(\sigma)t_b^{\mu\nu}(
\sigma')$, integrating (\ref{zbgen}) by parts over $\sigma\in\sheet_b$ and
using the identities (\ref{areaeltid}) we arrive at the final result for the
first $b=b'$ contributions to (\ref{sebbp}),
\beq
S_E({\cal E}_b)=\mu_b\int_{\sheet_b}d^2\sigma~\sqrt{\eta_b(\sigma)}
+{1\over\alpha_b}\int_{\sheet_b}d^2\sigma~\sqrt{\eta_b(\sigma)}~\eta_b^{
\alpha\beta}(\sigma){\partial t_{b,\mu\nu}(\sigma)\over\partial\sigma
^\alpha}{\partial t_b^{\mu\nu}(\sigma)\over\partial\sigma^\beta}
\label{seb}\eeq
where
\beq
\mu_b={16\pi^3\over
k^2}\phi^2_b\lim_{\Lambda\to\infty}\frac{C_\Lambda}{\Lambda^2}~~~~~~,~~~~~~
\frac1{\alpha_b}=-\frac{4\pi^3}{k^2}\phi^2_b\lim_{\Lambda\to\infty}
\frac{C_\Lambda}{\Lambda^4}
\label{mualpha}\eeq
The first surface integral in (\ref{seb}) is the area ${\cal A}(\sheet_b)$ of
$\sheet_b$ in $\M4$, while the second integral can be integrated by parts and
written in terms of the extrinsic curvature of $\sheet_b$ using (\ref{embhess})
as
\beq
\int_{\sheet_b}d^2\sigma~\sqrt{\eta_b(\sigma)}~\eta_b^{\alpha\beta}(\sigma)
{\partial t_{b,\mu\nu}(\sigma)\over\partial\sigma^\alpha}{\partial
t_b^{\mu\nu}(\sigma)\over\partial\sigma^\beta}=\int_{\sheet_b}d^2\sigma~
\sqrt{\eta_b(\sigma)}~K_{b,\ell;\beta}^\alpha(\sigma)K_{b,\ell;\alpha}
^\beta(\sigma)\equiv4\pi\chi_E^{(1)}(N\sheet_b)
\label{sebcurv}\eeq
Thus the contributions $S_E({\cal E}_b)$ to the action term (\ref{sesum})
describe the extrinsic geometry of the embedded surfaces in $\M4$. Note that
the curvature term (\ref{sebcurv}), which is the Euler number of the normal
bundle $N\sheet_b$, can be written in the form
\beq
\chi_E^{(1)}(N\sheet_b)=\frac1{4\pi}\int_{\sheet_b}d^2\sigma~
\sqrt{\eta_b(\sigma)}~\eta_b^{\alpha\beta}(\sigma)D_{b,\alpha}n_{b,\ell}^\mu
\,D_{b,\beta}n_{b,\ell;\mu}
\label{rigstringaction}\eeq
where
\beq
D_{b,\alpha}n_{b,\ell}^\mu={\partial n_{b,\ell}^\mu\over\partial
\sigma^\alpha}+A_{b,\alpha;\ell\ell'}\,n_{b,\ell'}^\mu=-K^\beta_{b,
\ell;\alpha}{\partial X_b^\mu\over\partial\sigma^\beta}
\label{dalpha}\eeq
and $A_{b,\alpha;\ell\ell'}=n_{b,\ell}^\mu{\partial n_{b,\ell';\mu}
\over\partial\sigma^\alpha}$ is an $SO(2)$ connection of the normal bundle
$N\sheet_b$.

\subsection{Extrinsic Intersection Numbers}

The remaining contribution $S_E({\cal N}_b)$ to the self-interaction terms in
(\ref{sebbp}) come from the points
$\{\sigma,\sigma'\}\in\sheet_b(\sigma)\otimes\sheet_b(\sigma')$ for which
$X_b^\mu(\sigma)=X_b^\mu(\sigma')$ but $\sigma\neq\sigma'$. The same structure
occurs for the $b\neq b'$ terms in (\ref{sebbp}), for which only $\cal N$-type
points contribute. We can therefore write the delta-function appearing in
(\ref{sebbp}) in terms of a delta-function on the space
$\sheet_b(\sigma)\otimes\sheet_{b'}(\sigma')$ to get
\beq\new{\begin{array}{c}
S_E({\cal N}_{bb'})={4\pi^2\over
k^2}\phi_b\phi_{b'}\sum_{\{\sigma_i,\sigma_i'\}\in
{\cal N}_{bb'}}\int_{\sheet_b}d^2\sigma~\sqrt{\eta_b(\sigma)}\int_{\sheet_{b'}}
d^2\sigma'~\sqrt{\eta_{b'}(\sigma')}\\\times
\,t_{b,\mu\nu}(\sigma)t_{b'}^{\mu\nu}
(\sigma')\,{\delta^{(2)}(\sigma,\sigma_i)\delta^{(2)}(\sigma',\sigma_i')\over
|J_i(\sigma,\sigma')|}\end{array}}
\label{senbdelta}\eeq
where $(X_b\otimes X_{b'})({\cal N}_{bb'})=\sheet_b\cap\sheet_{b'}$ and the
string worldsheets intersect transversally in $\M4$ at finitely many isolated
points $X_b^\mu(\sigma_i)=X_{b'}^\mu(\sigma_i')$. Here
$J_i(\sigma,\sigma')$ is the Jacobian for the four-dimensional coordinate
transformation $\{\sigma^\alpha,\sigma'^\alpha\}\to X_b^\mu(
\sigma)-X_{b'}^\mu(\sigma')$ on
$\sheet_b(\sigma)\otimes\sheet_{b'}(\sigma')\to\M4$. After a Taylor expansion
about the points $\sigma_i$ and $\sigma_i'$, we can work out this Jacobian at
the points $\sigma=\sigma_i$ and $\sigma'=\sigma_i'$ and we find
\beq
J_i(\sigma_i,\sigma_i')={1\over4}\left|\epsilon_{\mu\nu\lambda\rho}
\epsilon^{\alpha\beta}\epsilon^{\gamma\delta}{\partial X_b^\mu(\sigma_i)\over
\partial\sigma^\alpha}{\partial X_b^\nu(\sigma_i)\over\partial\sigma^\beta}
{\partial X_{b'}^\lambda(\sigma_i')\over\partial\sigma'^\gamma}{\partial
X_{b'}^\rho(\sigma_i')\over\partial\sigma'^\delta}\right|
\label{jacpts}\eeq
Substituting (\ref{jacpts}) into the action (\ref{senbdelta}) and integrating
over $\sigma$ and $\sigma'$ we have
\beq
S_E({\cal N}_{bb'})={16\pi^2\over
k^2}\phi_b\phi_{b'}\,\nu_G(\sheet_b,\sheet_{b'})
\label{intbbp}\eeq
where
\beq
\nu_G(\sheet_b,\sheet_{b'})=\sum_{\{\sigma,\sigma'\}\in{\cal N}_{bb'}}
{t_{b,\mu\nu}(\sigma)t_{b'}^{\mu\nu}(\sigma')\over\epsilon
^{\mu\nu\lambda\rho}t_{b,\mu\nu}(\sigma)t_{b',\lambda\rho}(\sigma')}
{}~{\rm sgn}\left(\epsilon^{\mu\nu\lambda\rho}t_{b,\mu\nu}(\sigma)t_{
b',\lambda\rho}(\sigma')\right)
\label{nugbbp}\eeq

The quantity (\ref{nugbbp}) is a geometrical intersection number of the
surfaces $\sheet_b$ and $\sheet_{b'}$ in $\M4$ (for $b=b'$ it is a
self-intersection number of $\sheet_b$). The sign function in (\ref{nugbbp}) is
the local intersection index of the intersection point
$p^\mu=X_b^\mu(\sigma)=X_{b'}^\mu(\sigma')$, and, since $t_{b,\mu\nu}$ is the
extrinsic area element, it takes the values $\pm1$ depending on whether or not
the orientation at $p$ coincides with that of $\M4$. The factor multiplying
each local intersection index is a geometrical quantity which measures the
transversality of the intersection. It vanishes as the normal vectors of
$\sheet_b$ at the intersection point $p$ become orthogonal, and it becomes
infinite as they become parallel. Thus the contributions to (\ref{sebbp}) from
the subsets ${\cal N}_{bb'}$ of points yields a signed geometric transversal
intersection index of the string worldsheets.

The intersection number (\ref{nugbbp}) can also be written in terms of
cohomology classes using the fact \cite{4man} that the deRham current of the
surface $\sheet_b$ can be written locally as
$(\triangle_{\sheet_b})_{\mu\nu}(x)=\int_{\sheet_b}d\sigma_{\mu\nu}(X_b)~
\delta^{(4)}(x,X_b(\sigma))$, so that
\beq
\nu_G(\sheet_b,\sheet_{b'})=\int_\M4\triangle_{\sheet_b}\wedge\star\,\triangle
_{\sheet_{b'}}=\int_{\sheet_b}\int_{\sheet_{b'}}\star\,\delta^{(2,2)}(X_b(
\sigma),X_{b'}(\sigma'))
\label{nugderham}\eeq
This shows that the geometrical intersection number is not a topological
invariant of the string worldsheets. When $b=b'$ and the $X_b$ are embeddings,
then (\ref{nugderham}) gives the first and second fundamental forms of the
embedded surface $\sheet_b$ in $\M4$ as described in the previous subsection.
This follows formally from the global property \cite{botttu}
\beq
\triangle_{\sheet_b}\wedge\triangle_{\sheet_{b'}}=\triangle_{\sheet_b\cap
\sheet_{b'}}\wedge\chi_E(N\sheet_b\cap N\sheet_{b'})
\label{globalprop}\eeq
of the deRham current, where $\chi_E$ denotes the Euler characteristic class.
If the geometry of the normal bundle of $\sheet_b$ is such that its deRham
current is self-dual, i.e. $\star\,\triangle_{\sheet_b}=\triangle_{\sheet_b}$,
then (\ref{nugderham}) coincides with the algebraic intersection number of
$\sheet_b$ and $\sheet_{b'}$ (see the next subsection). There exist examples of
K\"ahler 4-manifolds for which this is true \cite{4man}.

\subsection{Topological Intersection Numbers}

Finally, we come to the $\theta$-term in the effective action
(\ref{effaction}), which we can write as
\beq
S_I[\Sigma]\equiv-{4\pi\theta\over k^2}\int_\M4
d^4x~\epsilon_{\mu\nu\lambda\rho}\Sigma^
{\mu\nu}(x)\Sigma^{\lambda\rho}(x)=\sum_{b,b'}S_I^{(bb')}\label{si}\eeq
where
\beq\new{\begin{array}{lll}
S_I^{(bb')}&=&-{4\pi\theta\over k^2}\phi_b\phi_{b'}\int_{\sheet_b}d^2\sigma~
\sqrt{\eta_b(\sigma)}\int_{\sheet_{b'}}d^2\sigma'~\sqrt{\eta_{b'}(\sigma
')}\\& &~~~~~~~~~~\times\,
\epsilon^{\mu\nu\lambda\rho}t_{b,\mu\nu}(\sigma)t_{b,\lambda
\rho}(\sigma')\,\delta^{(4)}(X_{b'}(\sigma'),X_b(\sigma))\end{array}}
\label{sibbp}\eeq
The calculation proceeds as before. For the $b=b'$ terms in (\ref{sibbp}), the
contribution from the points $\{\sigma,\sigma'\}\in{\cal E}_b$ are calculated
using (\ref{zbgen}) with ${\cal K}(\sigma,\sigma')=\epsilon^{\mu\nu\lambda
\rho}t_{b,\mu\nu}(\sigma)t_{b,\lambda\rho}(\sigma')$. Because of the
identities (\ref{areaeltid}), the first term in (\ref{zbgen}) is absent in this
case, and integrating by parts over $\sheet_b$ we find
\beq
S_I({\cal E}_b)=-{16\pi^3\theta\over
k^2}\phi^2_b\left(\lim_{\Lambda\to\infty}\frac{C_\Lambda}{\Lambda^4}\right)
\nu_P(\sheet_b)\equiv\theta_b\,\nu_P(\sheet_b)
\label{sinup}\eeq
where
\beq
\nu_P(\sheet_b)={1\over4\pi}\int_{\sheet_b}d^2\sigma~\sqrt{\eta_b(
\sigma)}\,\eta_b^{\alpha\beta}(\sigma)\,\epsilon^{\mu\nu\lambda\rho}{\partial
t_{b,\mu\nu}(\sigma)\over\partial\sigma^\alpha}{\partial t_{b,\lambda\rho}
(\sigma)\over\partial\sigma^\beta}
\label{polindex}\eeq
is the self-intersection index of the worldsheet $\sheet_b$
\cite{whitney}--\cite{pawel}. It is the algebraic signed self-intersection
number of the surface, and it can be related to the Chern number of the normal
bundle of $\sheet_b$ by noting that from (\ref{fundform1}), (\ref{embhess}) and
(\ref{areaeltid}) we can write (\ref{polindex}) as
\beq
\nu_P(\sheet_b)={1\over8\pi}\int_{\sheet_b}d^2\sigma~\eta_b^{\alpha\beta}
\,\epsilon^{\gamma\delta}K_{b,\gamma\alpha}^\ell(\sigma)K_{b,\delta\beta}
^{\ell'}(\sigma)\,\epsilon_{\ell\ell'}={1\over2}\,{\rm ch}^{(1)}(N\sheet_b)
\label{polchern}\eeq
where
\beq
{\rm ch}^{(1)}(N\sheet_b)={1\over2\pi}\int_{\sheet_b}{\rm
tr}\,F_b={1\over8\pi}\int_
{\sheet_b}d^2\sigma~\epsilon_{\ell\ell'}\,\epsilon^{\alpha\beta}
F_{b,\alpha\beta}^{\ell\ell'}(\sigma)\label{chern1}\eeq
is the Chern number of $N\sheet_b$ and $F_{b,\ell\ell'}=dA_{b,\ell\ell'}$ is
the curvature of the $SO(2)$ connection of the normal bundle defined in
(\ref{dalpha}). Thus the self-intersection index also measures the
nontriviality of the normal bundle of $\sheet_b$ in $\M4$, and algebraically it
is counted by the Chern number of the normals.

Next we evaluate the contribution from $S_I({\cal N}_{bb'})$. Following the
steps which led to the expression (\ref{intbbp}), we find that it can be
written as
\beq
S_I({\cal N}_{bb'})=-{16\pi\theta\over k^2}\phi_b\phi_{b'}\,\nu_T(\sheet_b,
\sheet_{b'})
\label{intsibbp}\eeq
where
\beq
\nu_T(\sheet_b,\sheet_{b'})=\sum_{\{\sigma,\sigma'\}\in{\cal N}_{bb'}}
{\rm sgn}\left(\epsilon^{\mu\nu\lambda\rho}t_{b,\mu\nu}(\sigma)
t_{b,\lambda\rho}(\sigma')\right)
\label{nutbbp}\eeq
is a topological intersection number of the string worldsheets $\sheet_b$ and
$\sheet_{b'}$. As in (\ref{nugderham}) it can be expressed in terms of
cohomology classes as
\beq
\nu_T(\sheet_b,\sheet_{b'})=\int_\M4\triangle_{\sheet_b}\wedge
\triangle_{\sheet_{b'}}=\int_{\sheet_b}\int_{\sheet_{b'}}
\delta^{(2,2)}(X_b(\sigma),X_{b'}(\sigma'))
\label{nutderham}\eeq
showing that it is a topological invariant of the string worldsheets. The
quantity (\ref{nutderham}) is the algebraic intersection number of the oriented
surfaces $\sheet_b$ and $\sheet_{b'}$ \cite{4man}. When $b=b'$ and the $X_b$
are embeddings, (\ref{nutderham}) coincides with (\ref{polindex}) when the
geometry of the normal bundle of $\sheet_b$ is such that the Chern class of
$N\sheet_b$ coincides with the Poincar\'e class of $\sheet_b$. In the effective
field theory (\ref{effaction}), the instanton number of the complex line bundle
of the gauge theory, given by the topological Yang-Mills term in
(\ref{renaction}), is then the sum of the monopole numbers of the normal
bundles of the string worldsheets $\sheet_b$.

\subsection{Effective String Theory}

Collecting all of the contributions above, we find that the total effective
action (\ref{effaction}) of the deformed $BF$ field theory (\ref{renaction})
can be written in terms of geometrical and topological quantities as
\beq\new{\begin{array}{c}
\Gamma[\Sigma,j]=-{2\pi\over
k}\sum_{a,b}q_a\phi_b\,I(L_a,\sheet_b)+\sum_b\left(
\mu_b\,{\cal A}(\sheet_b)+{4\pi\over\alpha_b}\,\chi_E^{(1)}(N\sheet_b)
+\frac{\theta_b}2\,{\rm ch}^{(1)}(N\sheet_b)\right)\\+{16\pi\over
k^2}\sum_{b,b'}\phi_b\phi_{b'}\Bigl(\pi\,\nu_G(\sheet_b,
\sheet_{b'})-\theta\,\nu_T(\sheet_b,\sheet_{b'})\Bigr)\end{array}}
\label{effactiontot}\eeq
Each contribution to (\ref{effactiontot}) is a diffeomorphism invariant of the
embedded trajectories in $\M4$, as anticipated from the topological nature of
the field theory defined by (\ref{renaction}). The first sum in
(\ref{effactiontot}) shows that the composite particle-string states in the
spectrum of this quantum field theory have fractional exchange statistics. The
third sum yields analogous fractional geometrical and topological intersection
phases for the strings.

The second sum in (\ref{effactiontot}) has appeared in the context of the
extrinsic geometry of the QCD string \cite{polQCD,polstring,pawel}, and more
generally in the effective theory of Nielsen-Olesen vortex strings in abelian
Higgs field theories \cite{extcurv,topvortex}. The first term is the Nambu-Goto
area action while the second term is the rigid string action. The last term is
the analog of the $\theta$-term of four-dimensional Yang-Mills theory. It is
expected (from scale-invariance and loop equation arguments) that the correct
string theory for QCD is one in which the Nambu-Goto term is absent or
irrelevant, and the extrinsic curvature term controls the phase structure of
the string theory. It is interesting to note that this property is reflected by
the forms of the induced coefficients (\ref{mualpha}). When $\M4$ is an open
infinite spacetime, the string tension $\mu_b$ diverges, and to make the
effective string theory well-defined it should be set to 0. This can be
achieved in some limiting situtation involving the parameters $k$ and $\phi_b$
of the deformed $BF$ field theory. The only remnants of the action then are the
two topological terms, with a negative rigidity factor
\cite{extcurv,topvortex}. On the other hand, when $\M4$ is a compact manifold,
only the area form survives, consistent with the cohomological representations
(\ref{nugderham}) and (\ref{nutderham}). Notice that, for the special periodic
value $\theta=\pi$ of the vacuum angle, when the geometry of the string
worldsheets is such that their deRham currents are self-dual 2-forms, the sum
over intersection numbers in (\ref{effactiontot}) vanishes and we are left with
a pure effective string theory (plus an additional non-local particle-string
Aharonov-Bohm interaction \cite{ABphase,topvortex}). This is precisely the
value of $\theta$ that was found in \cite{pawel} to be induced by the dynamical
cancellation of folded string configurations. The deformed $BF$ field theory
(\ref{renaction}) is thus a dual model for rigid vortex strings with
$\theta$-vacua and additional non-local interaction terms \cite{topvortex}.
This topological field theory approach might serve as a useful tool for
investigating the physical properties of such systems. As we will see, the
topological nature of the dual model allows a complete and exact solution of
the quantum theory.

\section{Canonical Quantization of Deformed $BF$ Theory}

In the previous section we have uncovered a rich geometrical and topological
structure for the renormalized theory (\ref{renaction}) which has many
potential physical applications. We can learn more about this topological field
theory from canonical quantization. We will see that the quantization of it
will yield some novel quantum representations of the geometrical and
topological indices, just as the wavefunctions of ordinary $BF$ theory do for
the topological linking numbers and the cohomology of the underlying manifold.
In the following we shall be interested in precisely how these objects are
realized in the physical sector of the Hilbert space of this quantum field
theory. In this section we will describe the canonical structure of the field
theory (\ref{renaction}), taking into careful account the first-stage
reducibility of its gauge symmetries. The reduced phase space of similar
antisymmetric tensor field theories has been studied in \cite{lahiri}.

We choose the spacetime to be the product manifold $\M4={\Bbb R}\times\MT$,
where ${\Bbb R}$ parametrizes the time and $\MT$ is a 3-manifold without
boundary. We may then work in an adiabatic limit of the field theory in which
the temporal components of the particle and string source fields parametrize
their worldlines and worldsheets, i.e. $r_a^0(\tau)=\tau$ and
$X_b^0(\sigma^1,\sigma^2)=\sigma^1$. The temporal components $A_0$ and $B_{0i}$
are Lagrange multipliers which enforce the local gauge constraints
\beq
-\nabla_iF^{0i}+{k\over2\pi}\nabla\cdot B+j^0\approx0~~~~~~,~~~~~~{k\over4\pi}
\epsilon^{ijk}F_{jk}+\Sigma^{0i}\approx0\label{gaugeconstr}\eeq
where $B^i(x)={1\over2}\epsilon^{ijk}B_{jk}(x)$ and
$\epsilon^{ijk}\equiv\epsilon^{0ijk}$. From (\ref{gaugeconstr}) it follows
that, when $\MT$ is compact, $A$ and $B$ are only globally defined differential
forms on $\MT$ when the total particle charge and total string flux vanish,
$\sum_aq_a=\sum_b\phi_b=0$. When they are non-zero, the fields are instead
sections of a non-trivial complex line bundle over $\MT$ and the action
(\ref{renaction}) must be appropriately modified \cite{non0mod}. From a
physical point of view, the restriction to vanishing charge and flux sectors of
the theory on a closed space is natural by flux conservation. We shall assume
this constraint on the source currents in this paper. Some aspects of abelian
$BF$ theories on topologically non-trivial line bundles have been discussed
recently in \cite{cr}.

The canonical momenta in the temporal gauge $A_0=B_{0i}=0$ are
\beq
\pi^i\equiv\frac{\delta S}{\delta\dot
A_i}\biggm|_{j=\Sigma=0}=\dot{A}^i-{\theta\over\pi}\epsilon^{ijk}F_{jk}
{}~~~~~~,~~~~~~\pi^{ij}\equiv\frac{\delta S}{\delta\dot
B_{ij}}\biggm|_{j=\Sigma=0}={k\over2\pi}\epsilon^{ijk}A_k
\label{canmom}\eeq
and in this gauge we may invoke the strong equalities $\pi^0=\pi^{0i}=0$
\cite{lahiri}. They yield the non-vanishing canonical Poisson brackets
\beq
\left\{A_i(x),\pi^j(y)\right\}_P=\delta_i^j\,\delta^{(3)}(x,y)~~~~~~,~~~~~~
\left\{B_{ij}(x),\pi^{k\ell}(y)\right\}_P=\left(\delta_i^k\delta_j^\ell-
\delta_i^\ell\delta_j^k\right)\delta^{(3)}(x,y)\label{poisson}\eeq
and the canonical Hamiltonian
\beq
H=\int_\MT d^3x~\left[{1\over2}\left(\pi^i+{\theta\over\pi}\epsilon^{ijk}
F_{jk}\right)^2+{1\over4}F_{ij}F^{ij}-A_ij^i-{1\over2}B_{ij}\Sigma^{ij}\right]
\label{constrham}\eeq
We therefore have to quantize the constrained dynamical system with Hamiltonian
(\ref{constrham}), Poisson brackets (\ref{poisson}) and primary constraint
functions
\beq\new{\begin{array}{lll}
\lambda_1^{ij}(x)&=&\pi^{ij}(x)-{k\over2\pi}\epsilon^{ijk}A_k(x)\cr
\lambda_2(x)&=&\nabla\cdot\pi(x)-{k\over2\pi}\nabla\cdot B(x)-j^0(x)\cr
\lambda_3^i(x)&=&{k\over4\pi}\epsilon^{ijk}F_{jk}(x)+\Sigma^{0i}(x)
\end{array}}
\label{constrfuns}\eeq
The constraints $\lambda_a\approx0$ are first class constraints in the Dirac
classification of constrained systems \cite{dirac}, since they generate an
abelian Poisson-Lie algebra
\beq
\{\lambda_a,\lambda_b\}_P=0~~~~,~~a,b=1,2,3
\label{PLalg}\eeq

Secondary constraints are generated by the compatibility conditions for the
primary constraint functions (\ref{constrfuns}) given by
\beq
{\partial\over\partial
t}\lambda_a-\{\lambda_a,H\}_P+c_a^{~b}\,\lambda_b\approx0
\label{secconstr}\eeq
where $c_a^{~b}$ is a 2-cocycle of the abelian Poisson-Lie group generated by
the $\lambda_a$'s. This yields the additional constraint function
\beq
\lambda_4^i(x)=\pi^i(x)-{2\pi\over k}\epsilon^{ijk}\Sigma_{jk}(x)+{\theta
\over\pi}\epsilon^{ijk}F_{jk}(x)\label{l4}\eeq
It has vanishing Poisson bracket with itself and with $\lambda_2$, but
non-vanishing Poisson brackets with $\lambda_1$ and $\lambda_3$ in
(\ref{constrfuns}),
\beq
\left\{\lambda_1^{ij}(x),\lambda_4^k(y)\right\}_P={k\over2\pi}\epsilon^{ijk}
\,\delta^{(3)}(x,y)~~~~~~,~~~~~~\left\{\lambda_3^i(x),\lambda_4^j(y)\right\}_P
=-{k\over2\pi}\epsilon^{ijk}\nabla_k\,\delta^{(3)}(x,y)
\label{l134comm}\eeq
There are no tertiary constraints, owing to the first-stage reducibility of the
gauge theory, and (\ref{constrfuns}),(\ref{l4}) constitute the complete set of
constraint functions for the dynamical system.

We shall choose the pair $\lambda_1,\lambda_4$ as second class constraint
functions and impose the strong equalities
\beq
\pi^{ij}={k\over2\pi}\epsilon^{ijk}A_k~~~~~~,~~~~~~\pi^i={2\pi\over
k}\epsilon^{ijk}\Sigma_{jk}-{\theta\over\pi}\epsilon^{ijk}F_{jk}
\label{strongeq}\eeq
Then the remaining phase space variables have the non-vanishing Dirac brackets
\beq\new{\begin{array}{lll}
\left\{A_i(x),B^j(y)\right\}_D&\equiv&\left\{A_i(x),B^j(y)\right\}_P\\& &-
\int\!\!\!\int_\MT d^3x'~d^3y'~\Bigl\{A_i(x),\lambda_a(x')\Bigr\}_P\,({\cal
C}^{-1})^{ab}(x',y')\,\left\{\lambda_b(y'),B^j(y)\right\}_P\\&=&{2\pi\over
k}\,\delta^j_i\,\delta^{(3)}(x,y)\end{array}}
\label{diracbrac}\eeq
where ${\cal C}_{ab}=\{\lambda_a,\lambda_b\}_P$ is the Poisson bracket matrix
of the constraint functions for $a,b=1,4$. The constrained Hamiltonian is
\beq
H=\int_\MT d^3x~\left({4\pi^2\over k^2}\Sigma_{ij}\Sigma^{ij}+{1\over4}F_{ij}
F^{ij}-A_ij^i-{1\over2}B_{ij}\Sigma^{ij}\right)\label{ham}\eeq
with the additional first class constraints determined by the pair
$\lambda_2,\lambda_3$,
\beq
{4\pi\over k}\nabla\cdot\Sigma-{k\over2\pi}\nabla\cdot
B-j^0\approx0~~~~~~,~~~~~~
{k\over4\pi}\epsilon^{ijk}F_{jk}+\Sigma^{0i}\approx0\label{firstconstr}\eeq
where $\Sigma_i={1\over2}\epsilon_{ijk}\Sigma^{jk}$.

{}From the identity
\beq
{\partial\over\partial t}\nabla\cdot\Sigma=-{1\over2}\partial_\mu\left(
\epsilon^{\mu\nu\lambda\rho}\partial_\nu\Sigma_{\lambda\rho}\right)=0
\label{partid}\eeq
and those for $B$ and $F$ we see that the constraint functions in
(\ref{firstconstr}) are time independent. These constraints will be treated as
physical state conditions in the quantum field theory. From (\ref{diracbrac})
it follows that the non-vanishing canonical quantum commutators of the field
theory are
\beq
\left[A_i(x),B^j(y)\right]={2\pi i\over
k}\,\delta^j_i\,\delta^{(3)}(x,y)\label{quantcomm}\eeq
Notice that, in the absence of sources, the Hamiltonian (\ref{ham}) vanishes
only on the physical subspace of the entire Hilbert space. This again reflects
the ``mild" topological nature of deformed $BF$ theory, i.e. that the gauge
fermion field introduced by the deformation is constructed from the conjugate
momentum to the ghost field associated with the curvature constraint of the
field theory, as we discussed in section 2. In this sector, the reduced
classical phase space of the source-free field theory is the finite-dimensional
vector space
\beq
{\cal P}=H^1(\MT)\oplus H^2(\MT)
\label{topphasesp}\eeq
The physical Hilbert space therefore contains topological information and
yields quantum field theoretical representations of the deRham complex of the
3-manifold $\MT$, as is the usual case in topological gauge theories. This
space is studied in detail in the next section.

\section{Construction of the Physical Hilbert Space}

We now assume that $\MT$ is a compact, path-connected, orientable 3-manifold
without boundary, and let $p$ be the dimension of its first and second homology
groups\footnote{\baselineskip=12pt The ensuing construction also applies to the
case where $\MT$ is flat Euclidean 3-space. There $p=0$ and we assume that the
fields vanish at infinity.}. From the induced Euclidean-signature metric of
$\MT$ we can construct the dual forms $\tilde{j}$ and $\tilde{\Sigma}$ of the
vector fields (\ref{partcurr}) and (\ref{stringcurr}). The field $A$ restricted
to $\MT$ can be decomposed into exact, co-exact and harmonic forms using the
Hodge decomposition
\beq
A=d\vartheta+*dK'+a^\ell\alpha_\ell\label{ahodge}\eeq
where $\{\alpha_\ell\}_{\ell=1}^p\in H^1(\MT)$ is an orthonormal basis
of harmonic 1-forms and $*$ denotes the Hodge duality operator defined
with respect to the metric of $\MT$. The harmonic basis of 1-forms is chosen to
be Poincar\'e dual to a canonical homology basis of 2-cycles of $\MT$. Choosing
an orthonormal basis $\{\beta_\ell\}_{\ell=1}^p\in H^2(\MT)$ of harmonic
2-forms
in an analogous way, these generators have the normalizations
\beq
\int_\MT\alpha_\ell\wedge*\alpha_k=\int_\MT\beta_\ell\wedge*\beta_k=\delta_{\ell k}~~~~~~,~~~~~~\int_\MT\alpha_\ell\wedge\beta_k=M_{\ell k}\eeq
where $M_{\ell k}$ is the inverse of the topologically-invariant,
positive-definite integer-valued linking matrix $M^{k\ell}$ of the homology
1-cycles with the homology 2-cycles \cite{bss,botttu}. The scalar field
$\vartheta$, the 1-form field $K'$ and the harmonic coefficients $a_\ell$ are
formally given by
\beq\new{\begin{array}{c}
\nabla^2\vartheta=*d*A~~~~~~,~~~~~~d*dK'=F\\a^\ell(t)=M^{k\ell}\int_\MT A
\wedge\beta_k\end{array}}
\label{acoeffs}\eeq

Similarly, the Hodge decompositions of the 1-form fields
$*B$, $\tilde{j}$ and $*\tilde{\Sigma}$ over $\MT$ are
\beq\new{\begin{array}{c}
*B=d\vartheta'+*dK+b^\ell*\beta_\ell\\\nabla^2\vartheta'=*dB~~~~~~,~~~~~~
d*dK=d*B\\b^\ell(t)=M^{\ell k}\int_\MT B\wedge\alpha_k\end{array}}
\label{bcoeffs}\eeq

\beq\new{\begin{array}{c}
\tilde{j}=d\omega'+*d\Omega+j_\ell
M^{k\ell}*\beta_k\label{jhodge}\\\nabla^2\omega'=-{\partial j^0\over\partial
t}~~~~~~,~~~~~~d*d\Omega=d\tilde{j}\\
j_\ell(t)=\sum_aq_a\,\frac\partial{\partial
t}\left(\int_{r_0}^{r_a(t)}\alpha_\ell\right)\end{array}}
\label{jcoeffs}\eeq

\beq\new{\begin{array}{c}
*\tilde{\Sigma}=d\varpi+*d\Pi'+\Sigma_\ell M^{\ell
k}\alpha_k\\\nabla^2\varpi=*d\tilde\Sigma~~~~~~,~~~~~~*d*d\Pi'=-{\partial\Sigma_{0i}\over\partial t}\,dx^i\\
\Sigma_\ell(t)=\sum_b\phi_b\,\frac\partial{\partial
t}\left(\int_{\sheet_b(t)}\beta_\ell\right)\end{array}}
\label{sigcoeffs}\eeq
where $r_0$ is a fixed basepoint in $\MT$ and the surface $\sheet_b(t)$
represents the string worldsheet projected onto $\MT$ with boundary the string
$X_b(t,\sigma)$ at time $t$. In (\ref{jcoeffs}) and (\ref{sigcoeffs}) we have
used the continuity equations $\partial_\mu
j^\mu=\partial_\mu\Sigma^{\mu\nu}=0$ and the explicit forms (\ref{partcurr})
and (\ref{stringcurr}) of the sources.

It is convenient to introduce a holomorphic polarization for the harmonic
components of the gauge fields \cite{bss}. Consider the $2p$-dimensional phase
space (\ref{topphasesp}) of harmonic forms which represents the topological
degrees of freedom of the gauge fields that remain when there are no sources
present. On this space we introduce a complex structure defined by a symmetric
$p\times p$ complex-valued matrix $\rho$ such that $-\rho$ is an element of the
Siegal upper half-plane. Its imaginary part defines a metric
\beq
G^{\ell k}=-2M^{p\ell}~{\rm Im}\,\rho_{pq}~M^{qk}
\label{phasemetric}\eeq
on $\cal P$ and the desired polarization is defined by the complex variables
\beq
\gamma^\ell=a^\ell+M^{m\ell}\rho_{mk}b^k~~~~~~,~~~~~~\bar{\gamma}^\ell=a^\ell+
M^{m\ell}\bar{\rho}_{mk}b^k\label{gammaell}\eeq
In terms of the above decompositions, we find that the canonical quantum
commutator (\ref{quantcomm}) can be represented by the derivative operators
\beq\new{\begin{array}{c}
\vartheta'(x)={2\pi i\over k}{1\over\nabla^2_\perp}{\delta\over\delta
\vartheta(x)}~~~~~~,~~~~~~*F_i(x)={2\pi i\over k}P_{ij}{\delta\over\delta
K_j(x)}\\\bar{\gamma}^\ell={2\pi\over k}G^{\ell k}{\partial\over\partial
\gamma^k}\end{array}}
\label{derivops}\eeq
where $P_{ij}$ is the symmetric transverse projection operator on
$\Lambda^1(\MT)$ defined by
\beq
\nabla^iP_{ij}=\nabla^jP_{ij}=0~~~~~~,~~~~~~P_{ij}A^j=A_i-\nabla_i
\left({1\over\nabla^2_\perp}\nabla\cdot A\right)\eeq
and $\nabla_\perp^2$ denotes the scalar Laplacian with its zero modes removed.
The projections onto the subspaces orthogonal to the zero modes can be achieved
using time-independent gauge transformations and the vanishing condition on the
total flux of the sources.

Substituting (\ref{ahodge})--(\ref{sigcoeffs}) and (\ref{derivops}) into
(\ref{ham}) and integrating by parts, we find that the quantum Hamiltonian can
be decomposed into two commuting pieces as $H=H_L+H_T$. The local Hamiltonian
$H_L$ depends only on the local parts of the fields,
\beq\new{\begin{array}{c}
H_L=\int_\MT d^3x~\left[\left(-\vartheta{\partial j^0\over
\partial t}+{2\pi i\over k}\varpi(x){\delta\over\delta\vartheta}\right)+
\left(-K_i{\partial\Sigma^{0i}\over\partial t}-{2\pi i\over k}\Omega_iP^i_j
{\delta\over\delta K_j}\right)\right.\\\left.+{4\pi^2\over
k^2}\left(\varpi(x)\nabla^2
\varpi(x)+\Pi_i'(\nabla_1^2\Pi')^i-{1\over4}P_{ij}{\delta\over\delta K_j}P^i_k
{\delta\over\delta K_k}\right)\right]\end{array}}\label{hamell}\eeq
where $\nabla_1^2=*d*d$ is the Laplacian acting on co-exact 1-forms. The
topological Hamiltonian $H_T$ depends only on the global harmonic parts of the
fields,
\beq
H_T={4\pi^2\over k^2}M^{k\ell}M^{m\ell}\Sigma_k\Sigma_m+i(\Sigma_m-\bar{
\rho}_{mn}M^{n\ell}j_\ell)M^{mp}G_{pk}\gamma^k
-{2\pi i\over k}\left(\Sigma_k-\rho_{km}M^{m\ell}
j_\ell\right)M^{kn}{\partial\over\partial\gamma^n}\label{hamtop}\eeq
where $G_{\ell k}$ is the matrix inverse of $G^{\ell k}$.
In the Schr\"odinger picture, we can therefore separate the variables
$\vartheta$, $K$ and $\gamma$ and solve for the physical state wavefunctions
in the form
\beq
\Psi_{\rm phys}[\vartheta,K,\gamma;t]=\Psi_L[\vartheta,K;t]\,\Psi_T(\gamma;t)
\label{wavefuns}\eeq

\subsection{Local Gauge Symmetries and Adiabatic Linking Numbers}

The local wavefunctionals $\Psi_L$ must solve the first class constraints
(\ref{firstconstr}), which using the representations (\ref{derivops}) can be
written as
\beq\new{\begin{array}{rrl}
\left(i{\delta\over\delta\vartheta(x)}+j^0(x,t)-{4\pi\over k}\nabla\cdot
\Sigma(x,t)\right)\Psi_L[\vartheta,K;t]&=&0\\\left(iP^i_j{\delta\over\delta
K_j(x)}+\Sigma^{0i}(x,t)\right)\Psi_L[\vartheta,K;t]&=&0\end{array}}
\label{gaugederivs}\eeq
They are solved by wavefunctionals of the form
\beq
\Psi_L[\vartheta,K;t]=\exp\left[i\int_\MT d^3x~\left\{\vartheta(x)\left(
j^0(x,t)-{4\pi\over k}\nabla\cdot\Sigma(x,t)\right)+K_i(x)\Sigma^{0i}(x,t)
\right\}\right]\tilde{\Psi}_L(t)\label{gaugesol}\eeq
These wavefunctionals transform under the time-independent local gauge
transformations in (\ref{gaugetr}) which are exact,
\beq
A_i\to A_i+\nabla_i\chi'~~~~~~,~~~~~~B_{ij}\to B_{ij}+\nabla_i\xi_j'-
\nabla_j\xi_i'\label{localgaugetr}\eeq
This gauge symmetry is represented projectively in the states (\ref{gaugesol})
as
\beq\new{\begin{array}{l}
\Psi_L[\vartheta+\chi',K+\xi';t]\cr~~~~~~=\exp\left[i\int_\MT d^3x~\left\{\chi'
(x)\left(j^0(x,t)-{4\pi\over k}\nabla\cdot\Sigma(x,t)\right)+\xi_i'(x)
\Sigma^{0i}(x,t)\right\}\right]\Psi_L[\vartheta,K;t]\end{array}}
\label{gaugeprojrep}\eeq
in terms of a non-trivial local $U(1)\times U(1)$ 1-cocycle. The remaining
piece
$\tilde{\Psi}_L(t)$ is determined from the Schr\"odinger equation
\beq
i{\partial\over\partial t}\Psi_{\rm phys}[\vartheta,K,\gamma;t]=H
\Psi_{\rm phys}[\vartheta,K,\gamma;t]\label{schreq}\eeq
which using (\ref{hamell}) shows that it contains two contributions,
\beq
\tilde{\Psi}_L(t)=\e^{i\int_{-\infty}^tdt'~({\cal L}(t')+{\cal S}(t'))}
\label{psil}\eeq

In this subsection we shall discuss the first contribution to the local
wavefunctionals (\ref{psil}) which is the particle-string term
\beq
{\cal L}(t)={2\pi\over k}\int_\MT d^3x~\left(\varpi(x,t)j^0(x,t)-\Omega_i(x,t)
\Sigma^{0i}(x,t)\right)\label{calell}\eeq
This integral was evaluated in \cite{bss} using the relations (\ref{jcoeffs}),
(\ref{sigcoeffs}), (\ref{partcurr}), (\ref{stringcurr}) and shown to give
\beq
{\cal L}(t)=-{1\over2k}\sum_{a,b}q_a\phi_b\,{d\Phi_{ab}(t)\over dt}+{2
\pi\over k}j_\ell(t)M^{k\ell}\int_{-\infty}^tdt'~\Sigma_k(t')
\label{calellsum}\eeq
where
\beq\new{\begin{array}{c}
\Phi_{ab}(t)=4\pi\int_{-\infty}^t\left(\int_{{\mit\Sigma}_b(t')}\delta
^{(1,2)}\Bigl(r_a(t'),X_b(t',\sigma)\Bigr)\right)_i~dl^i(r_a(t'))\\
+\,4\pi\sum_{\lambda\neq0}{1\over\lambda^2}\psi_\lambda(r_a(t))\left(
\int_{{\mit\Sigma}_b(t)}*d\psi_\lambda\right)\end{array}}
\label{solidang}\eeq
is the generalized solid angle function on $\MT$. Here the Dirac delta-function
is defined as described in subsection 3.1 except now over $\MT$, and
$\psi_\lambda(x)$ are the orthonormal eigenfunctions of the scalar Laplacian
operator on $\MT$ with eigenvalue $\lambda^2$,
\beq
\nabla^2\psi_\lambda(x)=*d*d\psi_\lambda(x)=\lambda^2\psi_\lambda(x)
{}~~~~~~,~~~~~~\int_\MT\psi_\lambda*\psi_{\lambda'}=\delta_{\lambda\lambda'}
\label{eigenlapl}\eeq
Note that the Dirac delta-function $\delta^{(3)}(x,y)\in\Lambda^0(\MT)$ (or
$\delta^{(0,3)}(x,y)\in\Lambda^0(\MT(x))\otimes\Lambda^3(\MT(y))$) can then be
represented in terms of the completeness relation
\beq
\delta^{(3)}(x,y)=\sum_\lambda\psi_\lambda(x)\psi_\lambda(y)~~~~~~{\rm
or}~~\delta^{(0,3)}(x,y)\,d^3y=*\delta^{(3)}(x,y)=\sum_\lambda\psi_\lambda(x)
\otimes*\psi_\lambda(y)
\label{delta3reps}\eeq
If we further introduce a basis of orthonormal co-exact 1-forms
$\psi^{(c)}_{\tilde\lambda}$ which are the eigenstates of the Laplacian
operator $\nabla_1^2$ with eigenvalue $\tilde\lambda^2$,
\beq
\nabla_1^2
\psi_{\tilde\lambda}^{(c)}=*d*d\psi_{\tilde\lambda}^{(c)}=\tilde\lambda^2\psi_{
\tilde\lambda}^{(c)}~~~~~~,~~~~~~\int_\MT\psi_{\tilde\lambda}^{(c)}\wedge*
\psi_{\tilde\lambda'}^{(c)}=\delta_{\tilde\lambda\tilde\lambda'}
\label{co-exacteigen}\eeq
then the delta-function
$\delta^{(1,2)}(x,y)\in\Lambda^1(\MT(x))\otimes\Lambda^2(\MT(y))$ can be
represented in terms of the completeness relation \cite{bss}
\beq
\delta^{(1,2)}(x,y)=-\sum_{\lambda\neq0}\frac{d\psi_\lambda(x)\otimes*d\psi_
\lambda(y)}{\lambda^2}+\sum_{\tilde\lambda\neq0}\psi_{\tilde\lambda}^{(c)}(x)
\otimes*\psi_{\tilde\lambda}^{(c)}(y)+\alpha_\ell(x)\otimes M^{m\ell}\beta_m(y)
\label{delta12rep}\eeq

The function (\ref{solidang}) depends only on the topological classes of the
particle and string trajectories in $\MT$ and it represents the solid angle
formed by a string along $X_b(t,\sigma)$ relative to a particle at $r_a(t)$
\cite{bss}. It has the property that it changes by $4\pi$ everytime that a
particle is adiabatically transported around a fixed string (for which
$\sheet_b(t)$ is constant and only the first term in (\ref{solidang})
contributes), or a string around a fixed particle in the opposite direction
(for which $r_a(t)$ is constant and only the second term in (\ref{solidang})
contributes), as long as these trajectories do not intersect. $\Phi_{ab}(t)$ is
the multivalued angle function that one anticipates in a theory of adiabatic
transports, and we see that the first term in (\ref{psil}) represents the
non-trivial particle-string linkings. It is that part of the full wavefunction
that represents the exotic exchange holonomies between particles and strings
and is easily seen to be the adiabatic limit of the covariant linking number in
(\ref{linknum}) that arises in the effective field theory. When $\MT={\Bbb
R}^3$ the function (\ref{solidang}) reduces to the usual form of a solid angle
\cite{bss}.

\subsection{Adiabatic Intersection Indices and Euler Numbers}

In this subsection we will evaluate the second contribution to (\ref{psil})
which is given by the string-string term
\beq
{\cal S}(t)=-{\pi^2\over k^2}\int_\MT d^3x~\left(-4\varpi(x,t)\nabla^2
\varpi(x,t)+4\Pi_i'(x,t)(\nabla_1^2\Pi')^i(x,t)+\Sigma_{0i}(x,t)
\Sigma^{0i}(x,t)\right)\label{caless}\eeq
For this, we use the completeness relations (\ref{delta3reps}) and
(\ref{delta12rep}) along with (\ref{sigcoeffs}) and (\ref{stringcurr}) to write
\beq\new{\begin{array}{rrl}
\Sigma_{0i}(x,t)\,dx^i&=&\sum_b\phi_b\sum_{\tilde\lambda\neq0}\psi^{(c)}_{\tilde
\lambda}(x)\left(\int_{\partial\sheet_b(t)}\psi_{\tilde\lambda}^{(c)}\right)
\\\varpi(x,t)&=&-\sum_b\phi_b\sum_{\lambda\neq0}\frac{\psi_\lambda(x)}
{\lambda^2}\,\frac\partial{\partial
t}\left(\int_{\sheet_b(t)}*d\psi_\lambda\right)\end{array}}
\label{Sigpirep}\eeq
The 1-form $\Pi'$ is then written in terms of these eigenstates using
(\ref{sigcoeffs}). Substituting these decompositions into (\ref{caless}) and
integrating by parts using (\ref{eigenlapl}) and (\ref{co-exacteigen}) we then
have
\beq\new{\begin{array}{lll}
{\cal
S}(t)&=&\frac{4\pi^2}{k^2}\sum_{b,b'}\phi_b\phi_{b'}\left[\sum_{\lambda\neq0}
\frac1{\lambda^2}\,\frac\partial{\partial
t}\left(\int_{\sheet_b(t)}*d\psi_\lambda\right)\frac\partial{\partial
t}\left(\int_{\sheet_{b'}(t)}*d\psi_\lambda\right)\right.\\&
&\left.-\sum_{\tilde\lambda\neq0}\frac\partial{\partial
t}\left(\int_{\sheet_b(t)}*\psi_{\tilde\lambda}^{(c)}\right)\frac\partial
{\partial
t}\left(\int_{\sheet_{b'}(t)}*\psi_{\tilde\lambda}^{(c)}\right)-\frac14
\sum_{\tilde\lambda\neq0}\tilde\lambda^2\left(\int_{\sheet_b(t)}*\psi_{\tilde
\lambda}^{(c)}\right)\left(\int_{\sheet_{b'}(t)}*\psi_{\tilde\lambda}^{(c)}
\right)\right]\end{array}}
\label{calesseigen}\eeq
Integrating the first two terms in (\ref{calesseigen}) by parts over time gives
\beq\new{\begin{array}{lll}
{\cal
S}(t)&=&\frac{2\pi^2}{k^2}\sum_{b,b'}\phi_b\phi_{b'}\left[\frac{\partial^2}
{\partial
t^2}\left\{\sum_{\lambda\neq0}\frac1{\lambda^2}\left(\int_{\sheet_b(t)}
*d\psi_\lambda\right)\left(\int_{\sheet_{b'}(t)}*d\psi_\lambda\right)\right.
\right.\\&
&\left.-\sum_{\tilde\lambda\neq0}\left(\int_{\sheet_b(t)}*\psi_{\tilde
\lambda}^{(c)}\right)\left(\int_{\sheet_{b'}(t)}*\psi_{\tilde\lambda}^{(c)}
\right)-\left(\int_{\sheet_b(t)}*\alpha_\ell\right)\left(\int_{\sheet_{b'}(t)}
M^{m\ell}\beta_m\right)\right\}\\& &+\frac{\partial^2}{\partial
t^2}\left\{\left(\int_{\sheet_b(t)}*\alpha_\ell\right)\left(\int_{\sheet_{b'}
(t)}M^{m\ell}\beta_m\right)\right\}\\&
&-\sum_{\lambda\neq0}\frac1{\lambda^2}\left\{\left(
\int_{\sheet_b(t)}*d\psi_\lambda\right)\frac{\partial^2}{\partial
t^2}\left(\int_{\sheet_{b'}(t)}*d\psi_\lambda\right)+\frac{\partial^2}{\partial
t^2}\left(\int_{\sheet_b(t)}*d\psi_\lambda\right)\left(\int_{\sheet_{b'}(t)}
*d\psi_\lambda\right)\right\}\\&
&+\sum_{\tilde\lambda\neq0}\left\{\left(\int_{\sheet_b(t)}*\psi_{\tilde
\lambda}^{(c)}\right)\frac{\partial^2}{\partial
t^2}\left(\int_{\sheet_{b'}(t)}*\psi_{\tilde\lambda}^{(c)}\right)
+\frac{\partial^2}{\partial t^2}\left(\int_{\sheet_b(t)}*\psi_{\tilde
\lambda}^{(c)}\right)\left(\int_{\sheet_{b'}(t)}*\psi_{\tilde\lambda}^{(c)}
\right)\right\}\\&
&\left.-\frac14\sum_{\tilde\lambda\neq0}\left\{\left(\int_{\partial\sheet_b(t)}
*d\psi_{\tilde\lambda}^{(c)}\right)\left(\int_{\sheet_{b'}(t)}*\psi_{\tilde
\lambda}^{(c)}\right)+\left(\int_{\sheet_b(t)}*\psi_{\tilde
\lambda}^{(c)}\right)\left(\int_{\partial
\sheet_{b'}(t)}*d\psi_{\tilde\lambda}^{(c)}\right)\right\}\right]\end{array}}
\label{calessintparts}\eeq

The first three terms in (\ref{calessintparts}) can be combined together to
give the Dirac delta-function
$*\delta^{(1,2)}(x,x')\in\Lambda^2(\MT(x))\otimes\Lambda^2(\MT(x'))$ with
$x(t)\in\sheet_b(t)$ and $x'(t)\in\sheet_{b'}(t)$. The fourth term can be
rewritten using the harmonic coefficients $\Sigma_\ell(t)$ in (\ref{sigcoeffs})
and Hodge duality to relate the harmonic 1-forms and 2-forms by
$*\alpha_\ell=M^{k\ell}\beta_k$. In this way we arrive finally at the
expression
\beq
{\cal
S}(t)=-\frac1{4k^2}\sum_{b,b'}\phi_b\phi_{b'}\,\frac{d\Upsilon_{bb'}(t)}{dt}+
\frac{4\pi^2}{k^2}M^{k\ell}M^{m\ell}\,\frac
d{dt}\int_{-\infty}^tdt'~\Sigma_k(t)\Sigma_m(t')
\label{calessfinal}\eeq
where
\beq
\Upsilon_{bb'}(t)=8\pi^2\,\frac
d{dt}\int_{\sheet_b(t)}\int_{\sheet_{b'}(t)}*\,\delta^{(1,2)}
\Bigl(X_b(t,\sigma),X_{b'}(t,\sigma')\Bigr)+8\pi^2\,\Xi_E[N\sheet_b(t)\cap
N\sheet_{b'}(t)]
\label{adintindex}\eeq
is the generalized intersection number of the projected surfaces $\sheet_b(t)$
and $\sheet_{b'}(t)$ onto $\MT$ with
\beq\new{\begin{array}{l}
\Xi_E[N\sheet_b(t)\cap N\sheet_{b'}(t)]=\int_{-\infty}^tdt'~\left[\sum_{\lambda
\neq0}\frac1{\lambda^2}
\left\{\left(\int_{\sheet_b(t')}*d\psi_\lambda\right)\frac{d^2}{dt'^2}
\left(\int_{\sheet_{b'}(t')}*d\psi_\lambda\right)\right.\right.\\
\left.~~~~~~~~~~~~~~~~~~~~~~~~~~~~~~~~~~~~~~~~~~~~~~~~~~~~~~~+\frac{d^2}{dt'^2}
\left(\int_{\sheet_b(t')}*d\psi_\lambda\right)\left(\int_{\sheet_{b'}(t')}
*d\psi_\lambda\right)\right\}\\~~~~~~~~~~
-\sum_{\tilde\lambda\neq0}\left\{\left(\int_{\sheet_b(t')}*\psi_{\tilde
\lambda}^{(c)}\right)\frac{d^2}{dt'^2}\left(\int_{\sheet_{b'}(t')}*\psi_{\tilde
\lambda}^{(c)}\right)+\frac{d^2}{dt'^2}\left(\int_{\sheet_b(t')}*\psi_{\tilde
\lambda}^{(c)}\right)\left(\int_{\sheet_{b'}(t')}*\psi_{\tilde\lambda}^{(c)}
\right)\right\}\\~~~~~~~~~~
\left.+\frac14\sum_{\tilde\lambda\neq0}\left\{\left(\int_{\partial\sheet_b(t')}
*d\psi_{\tilde\lambda}^{(c)}\right)\left(\int_{\sheet_{b'}(t')}*\psi_{\tilde
\lambda}^{(c)}\right)+\left(\int_{\sheet_b(t')}*\psi_{\tilde
\lambda}^{(c)}\right)\left(\int_{\partial
\sheet_{b'}(t')}*d\psi_{\tilde\lambda}^{(c)}\right)\right\}\right]\end{array}}
\label{Eulerad}\eeq

Unlike the solid angle function (\ref{solidang}), this intersection function is
not a homological invariant of the surfaces $\sheet_b(t)$. Under a
homologically trivial motion $\sheet_b(t)\to\sheet_b(t)+\partial B_b(t)$ of a
given string for some volume $B_b(t)\subset\MT$, using Stokes' theorem we find
that the function (\ref{adintindex}) for $b\neq b'$ changes by
\beq
\delta\Upsilon_{bb'}(t)=-16\pi^2\int_{-\infty}^tdt'~\sum_{\lambda\neq0}\,\frac
d{dt'}\left(\int_{B_b(t')}\psi_\lambda(x)~d^3x\right)\frac{d}{dt'}\left(\int_{
\sheet_{b'}(t')}*d\psi_\lambda\right)
\label{homchangeups}\eeq
which is non-vanishing in general. It is, however, a smooth invariant of the
projected worldsheets. This is anticipated since (\ref{adintindex}) represents
the adiabatic limit of the extrinsic intersection index in $\M4$ which we
studied in section 3. If string $b'$ is held fixed and string $b\neq b'$ is
adiabatically transported through space, then only the delta-function in
(\ref{adintindex}) contributes and it counts extrinsic intersections of
$\sheet_b(t)$ and $\sheet_{b'}(t)$. Each such intersection essentially
contributes $8\pi^2$ to the function $\Upsilon_{bb'}(t)$. The time derivative
acts to give an extrinsic variation in the normal direction to $\MT$ and it is
the adiabatic analog of the transversality factor in (\ref{nugbbp}). The
function (\ref{Eulerad}) in general acts to make the adiabatic intersection
index well-defined for $b=b'$. When $b=b'$ and the surfaces do not
self-intersect, then the delta-function term in (\ref{adintindex}) and the last
sum in (\ref{Eulerad}) represent the adiabatic area form of $\sheet_b(t)$ while
the remaining terms in (\ref{Eulerad}) are the adiabatic limits of its
extrinsic curvature form.

Thus the second term in (\ref{psil}) represents a non-topological string-string
holonomy term which takes into account the intersections of the strings and
also the extrinsic geometry of their worldsheets. The function
$\Xi_E[N\sheet_b(t)\cap N\sheet_{b'}(t)]$ is the adiabatic limit of the Euler
numbers, which provide global corrections to the local intersection indices in
terms of the geometry and topology of the projected normal bundles of the
string worldsheets in ${\Bbb R}\times\MT$ (see (\ref{rigstringaction}) and
(\ref{globalprop})), and also of the transversality factor in (\ref{nugbbp}).
This follows from the way that it is explicitly related to the spectrum of the
Laplace-Beltrami operator of $\MT$ and that the one-forms $d\psi_\lambda$,
$\psi_{\tilde\lambda}^{(c)}$ and $*d\psi_{\tilde\lambda}^{(c)}$ can be regarded
as connections on $N\sheet_b(t)$. $\Upsilon_{bb'}(t)$ yields another sort of
multivalued ``angle function" in the wavefunctions of the deformed $BF$ theory
which is very different from the usual surface-surface linking
terms\footnote{\baselineskip=12pt For example, these intersection indices are
different than those found in \cite{lr} where the coupling of dynamical point
particles to $BF$ gauge fields was considered. The classical observables in
this case are related to mod 2 intersection numbers of $p$-chains and
$(d-p)$-chains, as well as the construction of global step functions, on
$d$-dimensional manifolds.}. Note that it is not possible to generically write
down an adiabatic limit of the topological intersection numbers which we
encountered in subsection 3.4, so that the canonical quantization procedure of
section 4 has consistently removed such potential topological string terms from
the Hamiltonian.

\subsection{Global Gauge Symmetries and Cohomology Representations}

The topological wavefunctions in (\ref{wavefuns}) represent the windings of the
sources around non-contractible cycles in $\MT$. From (\ref{hamtop}) and the
Schr\"odinger equation (\ref{schreq}) we see that they have the form
\beq\new{\begin{array}{lll}
\Psi_T(\gamma;t)&=&\exp\biggl[\int_{-\infty}^tdt'~\left(\Sigma_m(t')
-\bar{\rho}_{mn}(t')M^{n\ell}j_\ell(t')\right)M^{mp}G_{pk}\gamma^k
\cr& &-{4i\pi^2\over k^2}M^{k\ell}M^{m\ell}\int_{-\infty}^tdt'~\Sigma_k(t')
\Sigma_m(t')-{2\pi\over k}\int_{-\infty}^tdt'~\left(\Sigma_m(t')-\rho_{mp}
M^{p\ell}j_\ell(t')\right)\cr& &\times\,M^{mq}
G_{qr}M^{kr}\int_{-\infty}^{t'}dt''~\Bigl(\Sigma_k(t'')-\bar{
\rho}_{ks}M^{sn}j_n(t'')\Bigr)\biggr]\Psi_0(\gamma;t)\end{array}}
\label{psit}\eeq
where the function $\Psi_0$ is a solution of the equation
\beq
{\partial\Psi_0(\gamma;t)\over\partial t}=-{2\pi\over k}\left(\Sigma_k-\rho_{k
m}M^{m\ell}j_\ell\right)M^{kn}{\partial\Psi_0(\gamma;t)\over\partial\gamma^n}
\label{psi0eq}\eeq
The equation (\ref{psi0eq}) is solved by any function of the form
\beq
\Psi_0(\gamma;t)=\Psi_0\left(\gamma^\ell-{2\pi\over k}M^{k\ell}\int_{-
\infty}^tdt'~\Bigl(\Sigma_k(t')-\rho_{kn}M^{nm}j_m(t')\Bigr)\right)
\label{psi0gen}\eeq
The function $\Psi_0$ is fixed by requiring that, in addition to the local
gauge invariance (\ref{firstconstr}), the theory also be invariant under large
gauge
transformations which are not connected to the identity in $\cal P$, i.e. those
forms in (\ref{gaugetr}) which are not exact and have non-trivial cohomological
parts. For a consistent quantum theory, this global gauge symmetry must be
restricted to those forms in (\ref{gaugetr}) which have integer-valued
cohomology \cite{bss}, i.e. $\int_L\chi$ and $\int_{\mit\Sigma}\xi$ are integer
multiples of $2\pi$. When there are no sources present the wavefunctions
$\Psi_0$ should coincide with the cohomological states which represent the
invariance of the quantum field theory under these winding transformations.

In the source-free case the local gauge constraints (\ref{gaugederivs}) imply
that the physical state wavefunctions are functions only of the global harmonic
variables $\gamma$. Moreover, the Hamiltonian then vanishes when acting on
these states and so they are also time independent. In terms of the holomorphic
polarization (\ref{gammaell}) the restricted global gauge transformations are
\beq
\gamma^\ell\to\gamma^\ell+2\pi(n^\ell+M^{m\ell}\rho_{mk}m^k)~~~~~~,~~~~~~
\bar{\gamma}^\ell\to\bar{\gamma}^\ell+2\pi(n^\ell+M^{m\ell}\bar{\rho}_{mk}m^k
)\label{globgauge}\eeq
where $n^\ell$ and $m^\ell$ are integers. The invariance of the physical
states under the winding transformations (\ref{globgauge}) has been studied in
detail in \cite{bss} for the case when the coefficient $k$ of the $BF$ term in
(\ref{renaction}) is of the form $k=Mk_1/k_2$, where $M>0$ is the
integer-valued determinant of the linking matrix and $k_1$ and $k_2$ are
positive integers with ${\rm gcd}(Mk_1,k_2)=1$. We shall henceforth consider
only these values of $k$. This invariance condition is then uniquely solved by
the $(Mk_1k_2)^p$ independent holomorphic functions \cite{bss}
\beq
\Psi_0^{(q)}\pmatrix{c\cr d\cr}(\gamma)=\e^{(k/4\pi)\gamma^\ell G_
{\ell k}\gamma^k}~\Theta\pmatrix{{c+q\over Mk_1k_2}\cr d\cr}\left
({Mk_1\over2\pi}M_{\ell k}\gamma^k\biggm|-Mk_1k_2\rho\right)
\label{psi0q}\eeq
where $q^\ell=1,2,\ldots,Mk_1k_2$ and $\Theta$ are the doubly semi-periodic
Jacobi theta-functions
\beq
\Theta\pmatrix{c\cr d\cr}(z|\Pi)=\sum_{\{n_\ell\}\in{\Bbb Z}^p}\exp\left[i\pi
(n^\ell+c^\ell)\Pi_{\ell k}(n^k+c^k)+2\pi i(n^\ell+c^\ell)(z_\ell+d_\ell)
\right]\label{jactheta}\eeq
which are well-defined and holomorphic for $c^\ell,d_\ell\in[0,1]$, $\{z
_\ell\}\in{\Bbb C}^p$ and $\Pi=[\Pi_{\ell k}]$ in the Siegal upper half-plane.

The wavefunctions (\ref{psi0q}) are orthogonal in the inner product on $\cal P$
determined by the canonical coherent state measure for the holomorphic
polarization (\ref{gammaell}). The basis of states (\ref{psi0q}) thereby
produces an orthonormal basis of the full physical Hilbert space. They are
well-defined functions on the reduced topological phase space
\beq
{\cal P}_{\rm red}=H^1(\MT)\oplus H^2(\MT)/\Gamma_G
\label{redphasesp}\eeq
where $\Gamma_G\cong{\Bbb Z}^p\oplus{\Bbb Z}^p$ is the torsion-free part of the
integer cohomology group $H^1(\MT;{\Bbb Z})\oplus H^2(\MT;{\Bbb Z})$. The free
parameters $c$ and $d$ appearing in (\ref{psi0q}) can then be fixed by choosing
a spin structure on the complex $p$-torus (\ref{redphasesp}). From the
transformation properties of the Jacobi theta-functions under the modular group
$Sp(2p,{\Bbb Z})$ (acting on the matrix $\rho$) it follows that the physical
observables of the quantum field theory are independent of the choice of phase
space complex structure, as expected from the topological properties of the
theory.

If we denote the unitary quantum operators that generate the large gauge
transformations (\ref{globgauge}) on the Hilbert space by $U(n,m)$, then the
wavefunctions (\ref{psi0q}) transform under them as
\beq
U(n,m)\Psi_0^{(q)}\pmatrix{c\cr d\cr}(\gamma)=\sum_{q'}\,[U(n,m)]_{qq'}\,
\Psi_0^{(q')}\pmatrix{c\cr d\cr}(\gamma)
\label{psi0transf}\eeq
where the unitary matrices
\beq
[U(n,m)]_{qq'}=\exp\left[{2\pi i\over k_2}\left(c^\ell M_{m\ell}n^m+d_\ell
m^\ell+q^\ell M_{m\ell}n^m\right)-i\pi
kn^mM_{m\ell}m^\ell\right]\delta_{q^\ell-k_1Mm^\ell,q'^k}
\label{unitary}\eeq
generate a $(k_2)^p$-dimensional projective representation of the group
$\Gamma_G$ of large gauge transformations. The projective phases here are
non-trivial global $U(1)^p\times U(1)^p$ 1-cocycles and are cyclic with period
$k_2$. The topological part of the full wavefunction therefore carries a
non-trivial multi-dimensional representation of the discrete group $\Gamma_G$
representing the windings around the non-trivial homology cycles of $\MT$. The
invariance of the physical state wavefunctions under these global gauge
symmetries partitions the Hilbert space into superselection sectors labelled by
the integer cohomology classes of $\MT$, and thus the quantum states of the
deformed $BF$ theory provide novel quantum representations of the cohomology
ring $H^1(\MT)\oplus H^2(\MT)$. When combined with the explicit time dependence
(\ref{psil}), we shall see in the next section that the full wavefunctional
also carries a multi-dimensional projective representation of the algebra dual
to the algebra of large gauge transformations.

\section{Transformation Properties of the Physical States}

We will now examine the properties of the basis of full physical wavefunctions
(\ref{wavefuns}), which is given by combining together all of the components
(\ref{gaugesol}), (\ref{psil}), (\ref{calellsum}), (\ref{calessfinal}),
(\ref{psit}), (\ref{psi0gen}) and (\ref{psi0q}). Using (\ref{partcurr}) and
(\ref{stringcurr}), after some algebra we arrive at the total wavefunction
\beq\new{\begin{array}{l}
\Psi_{\rm phys}^{(q)}\pmatrix{c\cr
d\cr}[\vartheta,K,\gamma;t]=\exp\left[i\sum_aq_a\,\vartheta(r_a(t))\right.
\\~~~~~~~~~~~~~~~\left.+\,i\sum_b\phi_b\int d\sigma~\frac{\partial
X_b^i(t,\sigma)}{\partial\sigma}\left(K_i(X_b(t,\sigma))-\frac{4\pi}k
\,\epsilon_{ijk}\frac{\partial X_b^j(t,\sigma)}{\partial
t}\,\nabla^k\vartheta(X_b(t,\sigma))\right)\right]\\~~~~~~\times\exp\left[-\frac i{2k}\sum_{a,b}q_a\phi_b\Bigl(\Phi_{ab}(t)-\Phi_{ab}(-\infty)\Bigr)-\frac i{4k^2}\sum_{b,b'}\phi_b\phi_{b'}\Bigl(\Upsilon_{bb'}(t)-\Upsilon_{bb'}
(-\infty)\Bigr)\right.\\~~~~~~~~~~~~~~~+\frac
k{4\pi}\gamma^kG_{k\ell}\gamma^\ell+\frac{2\pi
i}k\int_{-\infty}^tdt'~j_\ell(t')M^{k\ell}\int_{-\infty}^{t'}dt''~\Sigma_k(t'')
\\~~~~~~~~~~~~~~~-\,i\gamma^k\int_{-\infty}^tdt'~j_k(t')-\frac{\pi
i}k\int_{-\infty}^tdt'~j_k(t')M^{pk}\rho_{pr}M^{r\ell}\int_{-\infty}^tdt''~
j_\ell(t'')\\~~~~~~~~~~~~~~~
\left.+\frac{4i\pi^2}{k^2}\int_{-\infty}^tdt'~\Sigma_k(t')M^{k\ell}
M^{m\ell}\Bigl(\Sigma_m(t)-\Sigma_m(t')\Bigr)\right]\\~~~~~~\times\,\Theta
\pmatrix{\frac{c+q}{Mk_1k_2}\cr
d\cr}\left(\frac{Mk_1}{2\pi}M_{k\ell}\gamma^k-k_2\int_{-\infty}^tdt'~\Bigl(
\Sigma_\ell(t')-\rho_{\ell
m}M^{mn}j_n(t')\Bigr)\biggm|-Mk_1k_2\rho\right)\end{array}}
\label{totalwavefn}\eeq
where $q^\ell=1,2,\dots,Mk_1k_2$, $\ell=1,\dots,p$, $k=Mk_1/k_2$, and the
topological components $j_\ell(t)$ and $\Sigma_\ell(t)$ of the sources are
given in (\ref{jcoeffs}) and (\ref{sigcoeffs}). The wavefunctions
(\ref{totalwavefn}) span a finite-dimensional Hilbert space and reduce to the
wavefunctions of pure $BF$ theory in the large-$k$ limit (as does the effective
action (\ref{effaction})), as anticipated since the coupling constant of the
quantum field theory is $1/k$. They represent the (exact) one-loop
renormalization of the wavefunctions of the canonical quantum field theory
(\ref{BFaction}).

The first exponential in (\ref{totalwavefn}) represents the invariance of the
physical states under local gauge transformations. It determines a
one-dimensional projective representation of the local $U(1)\times U(1)$ gauge
group with 1-cocycle
\beq
\Delta[\chi',\xi']=\frac1{2\pi}\sum_aq_a\,\chi'(r_a(t))+\frac1{2\pi}\sum_b\phi_b
\left[\int_{\sheet_b(t)}*\xi'-\frac{4\pi}k\,\frac\partial{\partial
t}\left(\int_{\sheet_b(t)}*d\chi'\right)\right]
\label{local1cocycle}\eeq
where $(\chi',\xi')\in\Lambda^0(\MT)\oplus\Lambda^1(\MT)$. This cocycle mixes
the local 1-form (particle) and 2-form (string) gauge degrees of freedom in a
non-trivial way and thereby defines a twisted representation of the local gauge
group. This projective representation therefore differs significantly from
those of usual topological gauge theories. The mixing term in
(\ref{local1cocycle}) can be absorbed into a secondary gauge transformation
(\ref{secondgaugesym}), so that the physical states also naturally carry a
representation of the secondary gauge symmetry.

The next set of local contributions involving the solid angle functions
$\Phi_{ab}(t)$ represent the adiabatic topological linking numbers of the
particle and string trajectories in $\MT$ and it endows the particle-string
wavefunctions with fractional exchange statistics, i.e. when a particle of
charge $q_a$ and a string of flux $\phi_b$ are adiabatically rotated once
around one another, the wavefunction acquires the phase
\beq
\hat\sigma=\e^{-{2\pi i\over k}q_a\phi_b}
\label{adphase}\eeq
The wavefunctionals (\ref{totalwavefn}) therefore carry, in addition to the
local gauge symmetries, a one-dimensional unitary representation of the
subgroup of the motion group of $\MT$ \cite{bss,motiongp} (the fundamental
homotopy group of the particle-string quantum configuration space) consisting
of the particle-string exchange holonomies. If $\MT$ is homologically trivial
then this is the full motion group of the space. When $\MT$ has non-trivial
homology we will see that the full wavefunctions also carry a representation of
the other part of the motion group associated with the windings of the
particles and strings around the non-trivial homology cycles of $\MT$.
Likewise, the set of local terms involving the intersection functions
$\Upsilon_{bb'}(t)$ represent non-topological intersection numbers as well as
the extrinsic geometry of the strings. They yield non-trivial phase factors in
the wavefunctions arising from intersections and self-intersections of the
worldsheets. They also provide a novel representation of the adiabatic limit of
the Euler characteristic classes of the normal bundles of the string
worldsheets.

When the particle and string sources are fixed, the topological components of
the wavefunctions (\ref{totalwavefn}) carry a twisted projective representation
of the global gauge group $\Gamma_G$ with 1-cocycle
\beq
\Delta\pmatrix{c^\ell\cr
d_\ell\cr}\left(n^\ell,m^\ell;k,M\right)=\frac1{k_2}\left(c^\ell
M_{r\ell}n^r+d_\ell m^\ell+q^\ell
M_{r\ell}n^r-\frac{Mk_1}2n^rM_{r\ell}m^\ell\right)
\label{global1cocycle}\eeq
where $(n,m)\in\Gamma_G$. The remaining components in (\ref{totalwavefn}) then
yield a ``topological duality" between representations of $\Gamma_G$, when we
consider homologically non-trivial motions of the particles and strings.
Consider the source configuration whereby the strings are fixed and the
particles wind $t_k$ times, up to a time $\tilde t$, around the $k^{\rm th}$
homology 1-cycle of $\MT$,
\beq
\int_{-\infty}^{\tilde t}dt'~j_k(t')=t_k~~~~~~,~~~~~~\int_{-\infty}^{\tilde
t}dt'~\Sigma_k(t')=0
\label{windingtildet}\eeq
and then afterwards the particles are fixed and the strings wind $s_k$ times,
up to a time $t>\tilde t$, around the $k^{\rm th}$ homology 2-cycle of $\MT$,
\beq
\int_{\tilde t}^tdt'~j_k(t')=0~~~~~~,~~~~~~\int_{\tilde
t}^tdt'~\Sigma_k(t')=s_k
\label{windingt}\eeq
The holonomies which arise from these configurations are taken into account by
the functions $\Phi_{ab}(t)$ and $\Upsilon_{bb'}(t)$. Modulo these holonomies,
the wavefunctions (\ref{totalwavefn}) transform under these motions as
\beq
\Psi_{\rm phys}^{(q)}[\vartheta,K,\gamma;t]\to\sum_{q'}\left[\widetilde{U}(s,t)
\right]_{qq'}\Psi_{\rm phys}^{(q')}[\vartheta,K,\gamma;-\infty]
\label{dualtransf}\eeq
where the unitary matrices
\beq
\left[\widetilde{U}(s,t)\right]_{qq'}=\exp\left[\frac{2\pi
i}{k_1M}\left(d_kM^{k\ell}t_\ell-s_kc^k-s_kq^k\right)+\frac{2\pi
i}ks_kM^{k\ell}t_\ell\right]\delta_{q^k-k_2M^{k\ell}t_\ell,q'^k}
\label{dualunitary}\eeq
generate a $(k_1)^p$-dimensional projective representation of $\Gamma_G$. This
representation is dual to the one in (\ref{psi0transf}), in that the
corresponding projective phase can be considered as the dual 1-cocycle to
(\ref{global1cocycle}),
\beq
\widetilde{\Delta}\pmatrix{c^\ell\cr
d_\ell\cr}(s_\ell,t_\ell;k,M)=\Delta\pmatrix{M^{\ell k}d_k\cr M_{\ell
k}c^k\cr}\left(-M_{\ell k}m^k,-M_{\ell k}n^k;\mbox{$\frac1k$}\,,M^{-1}\right)
\label{dualcocycle}\eeq
A similar sort of topological duality has been exploited recently in
\cite{mirror} to provide a deformed topological field theory interpretation of
the phenomenon of mirror symmetry in string theory.

This duality acts as both an $S$-duality $k\leftrightarrow\frac1k$, relating
perturbative and non-perturbative regimes of the quantum field theory and
interchanging electric charges with string fluxes in its spectrum, and also as
a Poincar\'e-Hodge duality relating non-trivial integer cohomology classes and
homology classes. The corresponding algebra of the unitary operators
(\ref{dualunitary}) consists of those generators of the motion group of $\MT$
which are associated with windings of the particles and strings around the
non-contractible cycles of $\MT$. Combining them with the local generators
represented by the phases (\ref{adphase}) we obtain an intriguing
representation of the full motion group of $\MT$ which can be described as
follows. Let $\hat s_k^{(\ell)}=\delta_{k\ell}$ and $\hat
t_k^{(m)}=\delta_{km}$. Then the operators
\beq
\hat\alpha_\ell=\widetilde{U}\left(\hat
s^{(\ell)},0\right)~~~~~~,~~~~~~\hat\beta_m=\widetilde{U}\left(0,\hat
t^{(m)}\right)
\label{braidops}\eeq
are the generators of motions of the particles and strings associated with the
$\ell^{\rm th}$ homology 1-cycle and $m^{\rm th}$ homology 2-cycle,
respectively. The operators (\ref{braidops}) together with (\ref{adphase})
generate a $(k_1)^p$-dimensional representation of the group of motions of the
particle worldlines and string worldsheets in the 3-manifold $\MT$ with the
relations
\beq\new{\begin{array}{rrl}
\Bigl[\hat\alpha_\ell,\hat\alpha_m\Bigr]~=~\left[\hat\beta_\ell,\hat\beta_m
\right]&=&0~=~\Bigl[\hat\alpha_\ell,\hat\sigma\Bigr]~=~\left[\hat\beta_\ell,
\hat\sigma\right]\\\hat\alpha_\ell\,\hat\beta_m&=&\e^{\frac{2\pi i}kM^{\ell
m}}\,\hat\beta_m\,\hat\alpha_\ell\end{array}}
\label{motionrels}\eeq
where we have used (\ref{dualunitary}). This is an abelian representation of
the motion group which generalizes the presentation given in \cite{motiongp}
from ${\Bbb R}^3$ to an arbitrary compact closed 3-manifold $\MT$. Note that in
this representation the linking matrix of $\MT$ plays a crucial role.
$M^{k\ell}$ is the identity matrix typically only for 3-manifolds $\MT$ which
are product spaces. Thus the $BF$ field theory also provides a natural way of
defining the motion group of generic manifolds.

The wavefunctions (\ref{totalwavefn}) thus incorporate the topology of the
underlying 3-manifold $\MT$ (via their dependence on the linking matrix
$M^{k\ell}$) and of its motion group (via the representation
(\ref{motionrels})) in precisely the same way that the wavefunctions of
ordinary $BF$ field theory do. The only overall effect of the topological
perturbation to the $BF$ action is to incorporate a sort of non-topological
holonomy factor for the intersections of the strings which is represented in
the twisted local 1-cocycle (\ref{local1cocycle}) and the intersection function
$\Upsilon_{bb'}(t)$. The wavefunctions (\ref{totalwavefn}) do, however,
represent the particle and string degrees of freedom in more symmetric fashion
and naturally incorporate the first-stage reducible gauge symmetries of the
topological field theory.

The deformed $BF$ theory thus yields quantum field theoretical representations
for new sorts of smooth invariants of 3-manifolds. The quantum holonomies
induced by these string-string terms could be relevant to the physics of
systems which involve vortex strings
\cite{bal1}--\cite{axion},\cite{fracstring}--\cite{electro}. In abelian Higgs
field theories, where the charged particles are represented by dynamical scalar
fields, the structures described in section 3 emerge as the leading orders of a
large Higgs mass expansion. The present formalism, which involves non-dynamical
point particles, naturally incorporates the particle-string Aharonov-Bohm
phases \cite{ABphase,topvortex}, extrinsic curvature terms
\cite{extcurv,topvortex}, and similar long-range string intersection
interactions \cite{topvortex} that have been discussed extensively in Higgs
models. The emergence of smooth surface invariants in this topological field
theory is intriguing in light of recent work \cite{cc-rr} on observables in
non-abelian $BF$ theories which suggests that surface observables yield
possibly new invariants of immersed surfaces in 4-manifolds. In the case of
non-topological deformations of $BF$ theory, these observables may be relevant
to the quark confinement problem \cite{BFQCD}. It would be interesting to
generalize the topological deformation we have considered in this paper to the
case of higher-dimensional $BF$ theories and to see what sort of smooth
invariants arise in these cases. This appears to be difficult to do within a
general framework, as the given class of gauge-invariant marginal deformations
depends crucially on the dimension of the underlying spacetime manifold.

\section*{Acknowledgements}

The author thanks M. Bergeron and G. Semenoff for an initial collaboration on
this subject, and A. Momen and J. Pawelczyk for comments on the manuscript.
Part of this work was carried out during the IIASS Workshop {\it Noncommutative
Geometry and Fundamental Interactions II} in Vietri, Italy in March 1998. The
author thanks the organisors and participants of the Workshop for having
provided a stimulating environment in which to work. This work was supported in
part by the Particle Physics and Astronomy Research Council (U.K.).

\end{document}